\documentclass[aps,prd,superscriptaddress,showpacs,preprint]{revtex4}
\usepackage{graphicx, bm}
\usepackage[usenames]{color}
\usepackage{float}

\begin{document}

\draft
\title{Model-independent limits for anomalous triple gauge bosons $W^+W^-\gamma$ coupling at the CLIC }

\author{A. A. Billur\footnote{abillur@cumhuriyet.edu.tr}}
\affiliation{\small Deparment of Physics, Sivas Cumhuriyet University, 58140, Sivas, Turkey.\\}

\author{M. K\"{o}ksal\footnote{mkoksal@cumhuriyet.edu.tr}}
\affiliation{\small Deparment of Optical Engineering, Sivas Cumhuriyet University, 58140, Sivas, Turkey.\\}

\author{ A. Guti\'errez-Rodr\'{\i}guez\footnote{alexgu@fisica.uaz.edu.mx}}
\affiliation{\small Facultad de F\'{\i}sica, Universidad Aut\'onoma de Zacatecas\\
         Apartado Postal C-580, 98060 Zacatecas, M\'exico.\\}

\author{ M. A. Hern\'andez-Ru\'{\i}z\footnote{mahernan@uaz.edu.mx}}
\affiliation{\small Unidad Acad\'emica de Ciencias Qu\'{\i}micas, Universidad Aut\'onoma de Zacatecas\\
         Apartado Postal C-585, 98060 Zacatecas, M\'exico.\\}

\date{\today}

\begin{abstract}

We investigate the potential of the Compact Linear Collider (CLIC) to probe the anomalous $W^+W^-\gamma$ coupling in $\gamma\gamma$,
$\gamma\gamma^*$ and $\gamma^*\gamma^*$ collisions through the processes $\gamma\gamma\rightarrow W^+ W^-$,
$e^+ \gamma \rightarrow e^+ \gamma^{*} \gamma \rightarrow e^+ W^- W^+$ and $e^+ e^-\rightarrow e^+ \gamma^{*} \gamma^{*} e^-
\rightarrow e^+ W^- W^+ e^-$. We perform leptonic, semi-leptonic and hadronic decays channels for $W^+W^-$ production in the 
final state. Taking ${\cal L}=1\hspace{0.8mm}{\rm ab^{-1}}@ \sqrt{s}=0.380\hspace{0.8mm}{\rm TeV}$, ${\cal L}=2.5\hspace{0.8mm}{\rm ab^{-1}}@ \sqrt{s}=1.5\hspace{0.8mm}{\rm TeV}$ and ${\cal L}=5\hspace{0.8mm}{\rm ab^{-1}}@ \sqrt{s}=3\hspace{0.8mm}{\rm TeV}$, based on
future CLIC data, the limits for $\Delta\kappa_\gamma$ and $\lambda_\gamma$ might reach up to ${\cal O}(10^{-5}-10^{-4})$ level
in the most ideal with 5\hspace{0.8mm}${\rm ab^{-1}}$ set data, which shows a potential advantage compared to those from LHC,
Tevatron and LEP data. Thus, our results represent that the processes $\gamma\gamma\rightarrow W^+ W^-$,
$e^+ \gamma \rightarrow e^+ \gamma^{*} \gamma \rightarrow e^+ W^- W^+$ and $e^+ e^-\rightarrow e^+ \gamma^{*} \gamma^{*} e^-
\rightarrow e^+ W^- W^+ e^-$ at the CLIC are a very good prospect for probing the anomalous $W^+W^-\gamma$ couplings.

\end{abstract}


\vspace{5mm}

\maketitle

\section{Introduction}

The Compact Linear Collider (CLIC) \cite{CLIC1,CLIC2,CLIC3,CLIC4} is proposed to testing the Standard Model (SM) \cite{SM} in detail,
with an unprecedented precision and with clean environments, in comparison with the hadron colliders. Its aim is to research the energy
frontier, providing sensitivity to physics beyond the Standard Model (BSM). As part of this physics program, the couplings between known
particles are parameterized in a general form to quantify possible deviations from their SM values, as is the case of the anomalous Triple
Gauge Boson Couplings (aTGC) $W^+W^-\gamma$, $W^+W^-Z$ \cite{Baur0,Hagiwara,Hagiwara1,Kumar,Aranda}. The measurement of the couplings between
the neutral electroweak bosons $Z$, $\gamma$ and the charged boson $W^\pm$, are the first two measurements that were able to prove the non-Abelian
character of the electroweak part of the SM \cite{SM}. Therefore, processes that are sensitive to gauge bosons self-interactions are important
tools used to search for non-standard effects.

A number of authors have made important contributions to the subject \cite{Baur0,Hagiwara,Hagiwara1,Nachtmann,Sahin,Cakir,Seyed,Ari}.
This topic has acquired new relevance in the recent years. In Table I, we summary $95\%$ Confidence Level (C.L.) limits on the aTGC
$\Delta\kappa_\gamma$ and $\lambda_\gamma$ from ATLAS, CMS, CDF, D0, ALEP, DELPHI, L3 and OPAL Collaborations and the future lepton
colliders such as the ILC and the CEPC. See Refs. \cite{Atag,Atag1,Sahin1,Papavassiliou,Choudhury,Chapon} for other limits on
$\Delta\kappa_\gamma$ and $\lambda_\gamma$ in different contexts.

The experiments at the CLIC in its three  phases ${\cal L}=1\hspace{0.8mm}{\rm ab^{-1}}@ \sqrt{s}=0.380\hspace{0.8mm}{\rm TeV}$,
${\cal L}=2.5\hspace{0.8mm}{\rm ab^{-1}}@ \sqrt{s}=1.5\hspace{0.8mm}{\rm TeV}$ and ${\cal L}=5\hspace{0.8mm}{\rm ab^{-1}}@
\sqrt{s}=3\hspace{0.8mm}{\rm TeV}$ can produce charged weak bosons ($W^\pm$) pairs in $\gamma\gamma$, $\gamma\gamma^*$ and
$\gamma^*\gamma^*$ collisions through the processes $\gamma\gamma\rightarrow W^+ W^-$, $e^+ \gamma \rightarrow e^+ \gamma^{*}
\gamma \rightarrow e^+ W^- W^+$ and $e^+ e^-\rightarrow e^+ \gamma^{*} \gamma^{*} e^-\rightarrow e^+ W^- W^+ e^-$. With these
energies and luminosities the CLIC can successfully achieve good limits on the anomalous $W^+W^-\gamma$ coupling, because
this collider is best suited for precision measurements \cite{CLIC1,CLIC2,CLIC3,CLIC4}. One of the advantages of $\gamma\gamma$,
$\gamma\gamma^*$ and $\gamma^*\gamma^*$ collisions is that they can isolate $W^+W^-\gamma$ couplings from $W^+W^-Z$ couplings
unlike $e^+e^-$  collisions. Also, $W^+W^-\gamma$ vertex makes contributes to $\gamma\gamma$, $\gamma\gamma^*$ and $\gamma^*\gamma^*$
collisions. These collisions are a particularly important tool in searching $W^\pm$ electromagnetic interactions. The processes
$\gamma\gamma (\gamma\gamma^*, \gamma^*\gamma^*) \to W^+W^-$ include only interactions between the gauge bosons, causing more
apparent possible deviations from the expected value of SM \cite{Eboli}. In addition, these processes provide the best
opportunity to measure directly the $W^+W^-\gamma$ aTGC via $t$ and $u$ channel (see Figs. 1-3), and as we already mentioned
can be used to test the non-Abelian nature of the SM.

\begin{table}[!ht]
\caption{Experimental and phenomenological limits at $95\%$ C.L. on the aTGC $\Delta\kappa_\gamma$ and $\lambda_\gamma$
from the present and future colliders.}
\begin{center}
\begin{tabular}{|c|c|c|c|c|}
\hline\hline
{\bf Model}             &    {$\Delta\kappa_\gamma$}  &  {$\lambda_\gamma$}          & {\bf C. L.}   &  {\bf Reference}\\
\hline
SM                      &           0                 &    0                         &               &   \cite{SM}  \\
\hline
\hline
\hline
{\bf Experimental limit}      &  {$\Delta\kappa_\gamma$}    &  {$\lambda_\gamma$}          &  {\bf C. L.}  &  {\bf Reference}\\
\hline
ATLAS Collaboration              &    [-0.061, 0.064]          & [-0.013, 0.013]      & $95 \%$    & \cite{ATLAS} \\
\hline
CMS Collaboration                &    [-0.044, 0.063]          & [-0.011, 0.011]      & $95 \%$    & \cite{CMS} \\
\hline
CDF Collaboration                &    [-0.158, 0.255]          & [-0.034, 0.042]      & $95 \%$    & \cite{CDF} \\
\hline
D0 Collaboration                 &    [-0.158, 0.255]          & [-0.034, 0.042]      & $95 \%$    & \cite{D0} \\
\hline
ALEP, DELPHI, L3, OPAL           &    [-0.099, 0.066]          & [-0.059, 0.017]      & $95 \%$    & \cite{LEP} \\
\hline
\hline
\hline
{\bf Phenomenological limit}           &  {$\Delta\kappa_\gamma$}    &  {$\lambda_\gamma$}  &{\bf C. L.} &  {\bf Reference}\\
\hline
ILC                              &    [-0.00037, 0.00037]      & [-0.00051, 0.00051]  & $95 \%$    & \cite{ILC} \\
\hline
CEPC                             &    [-0.00045, 0.00045]      & [-0.00033, 0.00033]  & $95 \%$    & \cite{CEPC} \\
\hline\hline
\end{tabular}
\end{center}
\end{table}

The rest of the paper is organized as follows: In Section II, the gauge-invariant operators of dimension-six are given. In Section III
and IV, we study the aTGCs $\Delta\kappa_\gamma$ and $\lambda_\gamma$ through the processes $\gamma\gamma\rightarrow W^+ W^-$,
$e^+ \gamma \rightarrow e^+ \gamma^{*} \gamma \rightarrow e^+ W^- W^+$ and $e^+ e^-\rightarrow e^+ \gamma^{*} \gamma^{*} e^-
\rightarrow e^+ W^- W^+ e^-$ at the $\gamma \gamma$, $\gamma \gamma^*$ and $\gamma^* \gamma^*$ collision mode. Finally, we present our
conclusions in Section V.

\section{The triple gauge boson vertex $W^+W^-\gamma$ with the anomalous contribution}

In this paper, effective Lagrangian technique is used to understand potential deviations from the SM predictions.
We adopt the effective Lagrangian for $W^+W^-\gamma$ interaction of the photon and the gauge bosons with operators
up to mass dimension-six:

\begin{equation}
{\cal L}_{eff}={\cal L}^{(4)}_{SM} + \sum_i \frac{C^{(6)}_i}{\Lambda^2}{\cal O}^{(6)}_i + {\rm h.c.},
\end{equation}

\noindent where ${\cal L}^{(4)}_{SM}$ denotes the renormalizable SM Lagrangian and ${\cal O}^{(6)}_i$ are the gauge-invariant
operators of mass dimension-six. The index $i$ runs over all operators of the given mass dimension. The mass scale is set by
$\Lambda$, and the coefficients $C_i$ are dimensionless parameters, which are determined once the full theory is known.

The effective Lagrangian relevant to the analysis of aTGC reads:

\begin{equation}
{\cal L}_{eff}=\frac{1}{\Lambda^2}\Bigl[C_W{\cal O}_W + C_B{\cal O}_B + C_{WWW}{\cal O}_{WWW} + \mbox{h.c.}\Bigr],
\end{equation}

\noindent with

\begin{eqnarray}
{\cal O}_W&=&\bigl( D_\mu\Phi \bigr)^{\dagger}\hat W^{\mu\nu}\bigl( D_\nu\Phi \bigr),\\
{\cal O}_B&=&\bigl( D_\mu\Phi \bigr)^{\dagger}\hat B^{\mu\nu}\bigl( D_\nu\Phi \bigr),\\
{\cal O}_{WWW}&=&Tr\bigl[\hat W^{\mu\nu}\hat W^{\nu\rho}\hat W^{\mu}_\rho\bigr].
\end{eqnarray}

\noindent $D_\mu$ is the covariant derivative, $\Phi$ is the Higgs doublet field and $\hat B_{\mu\nu}$, and $\hat W_{\mu\nu}$
are the $U(1)_Y$ and $SU(2)_L$ gauge field strength tensors. The coefficients of these operators $C_W/\Lambda^2$, $C_B/\Lambda^2$,
and $C_{WWW}/\Lambda^2$, are zero in the SM.

Based on this methodology, the effective Lagrangian for describe the $W^+W^-\gamma$ coupling can be parameterized as \cite{,Hagiwara,Gaemers}:

\begin{equation}
{\cal L}_{WW\gamma}=-ig_{WW\gamma} \Bigl[g^\gamma_1(W^\dagger_{\mu\nu} W^\mu A^\nu - W^{\mu\nu} W^\dagger_\mu A_\nu )
+\kappa_\gamma W^\dagger_{\mu} W_\nu A^{\mu\nu} + \frac{\lambda_\gamma}{M^2_W} W^\dagger_{\rho\mu} W^\mu_\nu A^{\nu\rho} \Bigr],
\end{equation}

\noindent where $g_{WW\gamma}=e$, $V_{\mu\nu}=\partial_\mu V_\nu -\partial_\nu V_\mu$ with $V_\mu= W_\mu, A_\mu$. The couplings
$g^\gamma_1$, $\kappa_\gamma$ and $\lambda_\gamma$ are CP-preserving, and in the SM their values are $g^\gamma_1=\kappa_\gamma=1$
and $\lambda_\gamma=0$ at the tree level.

The three operators of dimension-six, given by Eq. (2) are related to the aTGC via \cite{Rujula,Hagiwara1,Hagiwara2}:

\begin{equation}
\kappa_\gamma= 1+ \Delta\kappa_\gamma,
\end{equation}

\noindent with

\begin{eqnarray}
\Delta\kappa_\gamma&=&C_W + C_B,\\
\lambda_\gamma&=&C_{WW}.
\end{eqnarray}

From the effective Lagrangian given by Eq. (6) which CP-conserving, the Feynman rule for the anomalous $W^+W^-\gamma$
vertex is given by \cite{Hagiwara}:

\begin{eqnarray}
\Gamma^{WW\gamma}_{\mu\nu\rho}&=&e\Bigl[g_{\mu\nu}(p_1-p_2)_\rho + g_{\nu\rho}(p_2-p_3)_\mu + g_{\rho\mu}(p_3-p_1)_\nu
+\Delta\kappa_\gamma \Bigl( g_{\rho\mu}p_{3\nu}- g_{\nu\rho}p_{3\mu}\Bigr )  \nonumber\\
&+&\frac{\lambda_\gamma}{M^2_W}\Bigl(p_{1\rho}p_{2\mu}p_{3\nu} - p_{1\nu}p_{2\rho}p_{3\mu}
- g_{\mu\nu}(p_2\cdot p_3 p_{1\rho}-p_3\cdot p_1 p_{2\rho}) \nonumber\\
&-& g_{\nu\rho}(p_3\cdot p_1 p_{2\mu}-p_1\cdot p_2 p_{3\mu})
- g_{\mu\rho}(p_1\cdot p_2 p_{3\nu}-p_2\cdot p_3 p_{1\nu}) \Bigr)\Bigr].
\end{eqnarray}

\noindent Here, $p_1$ represents the momentum of the photon and $p_2$ and $p_3$ represent the momenta of $W^\pm$ bosons.
In addition, the first three terms in Eq. (10) corresponds to the SM couplings, while the terms with $\Delta\kappa_\gamma$
and $\lambda_\gamma$ give rise to aTGC.

Several searches on these anomalous $W^+W^-\gamma$ couplings $\Delta\kappa_\gamma$ and $\lambda_\gamma$ were performed by the LEP,
Tevatron and LHC experiments, as shown in Table I.

\section{Cross-section and Model-independent limits on the anomalous couplings $\Delta\kappa_\gamma$ and $\lambda_\gamma$
in $\gamma\gamma$, $\gamma\gamma^*$ and $\gamma^*\gamma^*$ collisions at the CLIC}

The advantage of the linear $e^+e^-$ colliders with respect to the hadron colliders is in the general cleanliness of the events
where two elementary particles, electron and positron beams, collide at high energy, and the high resolutions of the detector
are made possible by the relatively low absolute rate of background events. Furthermore, 1) these colliders will complement the
physics program of the LHC, especially for precision measurements. 2) Photon colliders $\gamma\gamma$ and $\gamma e^-$ have been
considered a natural addition to $e^+e^-$ linear colliders as the ILC and the CLIC. 3) The photon
colliders based on the ILC or the CLIC are the most realistic project. 4) Currently, the ILC and the CLIC are the best place for the
photon collider. In the case of the processes studied in this paper $\gamma\gamma\rightarrow W^+ W^-$, $e^+ \gamma \rightarrow
e^+ \gamma^{*} \gamma \rightarrow e^+ W^- W^+$ and $e^+ e^-\rightarrow e^+ \gamma^{*} \gamma^{*} e^-\rightarrow e^+ W^- W^+ e^-$
where $\gamma$ and $\gamma^*$ are Compton backscattered and Weizs\"{a}cker-Williams photons, they are extremely clean reactions because
there is no interference with weak and strong interactions as they are purely quantum electroweak reactions. This is very
useful for any new physics study, in particular, to study the anomalous $W^+W^-\gamma$ coupling.

The CLIC is a multi-TeV  linear $e^+e^-$ collider designed to operate in several center-of-mass energy stages,
as shown in Table II. As we mentioned above, this allows the study of the $\gamma\gamma$ and $\gamma e^-$ interactions by
converting the original $e^-$ or $e^+$ beam into a photon beam through the Compton backscattering mechanism. Other well-known
applications of the linear colliders are the processes $e\gamma^{*}$, $\gamma \gamma^{*}$ and $\gamma^{*} \gamma^{*}$ where the
emitted quasi-real photon $\gamma^{*}$ is scattered with small angles from the beam pipe of $e^{-}$ or $e^{+}$ \cite{Ginzburg,Ginzburg1,Brodsky,Budnev,Terazawa,Yang}. Since these photons have a low virtuality, they are almost on the mass shell.
These processes can be described by the Weizsacker-Williams Approximation (WWA) \cite{Budnev,Chen,Baur}. The WWA has a lot of advantages
such as providing the skill to reach crude numerical predictions via simple formulae. In addition, it may principally ease the experimental
analysis because it enables one to directly achieve a rough cross section for $\gamma^{*} \gamma^{*}\rightarrow X$ process via the
examination of the main process $e^{-}e^{+}\rightarrow e^{-} X e^{+}$ where X represents objects produced in the final state. The production
of high mass objects is particularly interesting at the linear colliders and the production rate of massive objects is limited by the photon
luminosity at high invariant mass while $\gamma^{*}\gamma^{*}$ and $e\gamma^{*}$ processes at the linear colliders arise from quasi-real
photon emitted from the incoming beams. Hence, $\gamma^{*} \gamma^{*}$ and $e\gamma^{*}$ are more realistic than $\gamma\gamma$ and
$e\gamma$. These processes have been observed experimentally at the LEP, the Tevatron and the LHC \cite{Abulencia,Aaltonen1,Aaltonen2,Chatrchyan1,Chatrchyan2,Abazov,Chatrchyan3}.

\begin{table}
\caption{Benchmark parameters of the  CLIC  for each stage in the updated scenario \cite{CLIC1,CLIC2,CLIC3,CLIC4}.}
\label{tab:1}
\begin{tabular}{|c|c|c|}
\hline
\hline
CLIC           &  $\sqrt{s}\hspace{0.8mm}$(TeV)    &  ${\cal L}(\rm fb^{-1})$      \\
\hline\hline
Stage 1    & 0.380                             & 100, 300, 500, 700, 1000     \\
\hline
Stage 2   & 1.5                               & 100, 500, 1000, 1500, 2500     \\
\hline
Stage 3    & 3                               & 100, 1000, 3000, 4000, 5000     \\
\hline
\hline
\end{tabular}
\end{table}

\subsection{The total cross-section of the processes $\gamma\gamma\rightarrow W^+ W^-$,
$e^+ \gamma \rightarrow e^+ \gamma^{*} \gamma \rightarrow e^+ W^- W^+$ and $e^+ e^-\rightarrow e^+ \gamma^{*} \gamma^{*} e^-
\rightarrow e^+ W^- W^+ e^-$ at the CLIC}

The production processes of diboson $\gamma\gamma\rightarrow W^+ W^-$, $e^+ \gamma \rightarrow e^+ \gamma^{*} \gamma \rightarrow
e^+ W^- W^+$ and $e^+ e^-\rightarrow e^+ \gamma^{*} \gamma^{*} e^-\rightarrow e^+ W^- W^+ e^-$, can also be produced from the
radiative couplings to photon, as shown in Figs. 1-3.

In future lepton linear colliders, such as the CLIC, high luminosity photon beams can be obtained by Compton backscattering of
a low-energy, high-intensity laser beam off the high-energy electron beam, and then $\gamma\gamma \to W^- W^+$ can be produced
from the laser photon fusion processes as shown in Fig. 3. In this case, the spectrum of Compton backscattered photons \cite{Ginzburg,Telnov}
is given by:

\begin{eqnarray}
 f_{\gamma}(y)=\frac{1}{g(\zeta)}\Bigl[1-y+\frac{1}{1-y}-
 \frac{4y}{\zeta(1-y)}+\frac{4y^{2}}{\zeta^{2}(1-y)^{2}}\Bigr] ,
\end{eqnarray}

\noindent where

 \begin{eqnarray}
 g(\zeta)=\Bigl(1-\frac{4}{\zeta}-\frac{8}{\zeta^2}\Bigr)\log{(\zeta+1)}+
 \frac{1}{2}+\frac{8}{\zeta}-\frac{1}{2(\zeta+1)^2} ,
 \end{eqnarray}

\noindent with

 \begin{eqnarray}
 y=\frac{E_{\gamma}}{E_{e}} , \;\;\;\; \zeta=\frac{4E_{0}E_{e}}{M_{e}^2}
 ,\;\;\;\; y_{max}=\frac{\zeta}{1+\zeta}.
 \end{eqnarray}

Here, $y$ is the fraction of electron energy carried away by the scattered photon, $E_{0}$ and $E_{e}$ are  energy of the incoming
laser photon and initial energy of the electron beam before Compton backscattering and $E_{\gamma}$ is the energy of the backscattered
photon. The maximum value of $y_{max}=\frac{\zeta}{1+\zeta}$ reaches 0.83 when $\zeta=4.8$, that is when the photon conversion
efficiency drops drastically, as a consequence of the $e^+e^-$ pair production from the laser photons and the photon
backscattering.

Other implementations of the linear colliders are the processes $e^+ \gamma \rightarrow e^+ \gamma^{*} \gamma \rightarrow e^+ W^- W^+$
and $e^+ e^-\rightarrow e^+ \gamma^{*} \gamma^{*} e^- \rightarrow e^+ W^- W^+ e^-$ described by the WWA, where the spectrum of photons
emitted by electrons is given by \cite{Belyaev,Budnev}:

\begin{eqnarray}
f_{\gamma^{*}}(x_{1})=\frac{\alpha}{\pi E_{e}}\Bigl\{\Bigl[\frac{1-x_{1}+x_{1}^{2}/2}{x_{1}}\Bigr]
log \Bigl(\frac{Q_{max}^{2}}{Q_{min}^{2}}\Bigr)-\frac{m_{e}^{2}x_{1}}{Q_{min}^{2}}
&&\Bigl(1-\frac{Q_{min}^{2}}{Q_{max}^{2}}\Bigr)-\frac{1}{x_{1}}\Bigl[1-\frac{x_{1}}{2}\Bigr]^{2}
log\Bigl(\frac{x_{1}^{2}E_{e}^{2}+Q_{max}^{2}}{x_{1}^{2}E_{e}^{2}+Q_{min}^{2}}\Bigr)\Bigr\}, \nonumber \\
\end{eqnarray}

\noindent with $x_1=E_{\gamma^*_e}/E_{e}$ and $Q^2_{max}$ is maximum virtuality of the photon. The minimum value of the $Q^2_{min}$
is given by:

\begin{eqnarray}
Q_{min}^{2}=\frac{m_{e}^{2}x_{1}^{2}}{1-x_{1}}.
\end{eqnarray}

Therefore, the total cross-section of the reactions $\gamma\gamma\rightarrow W^+ W^-$,
$e^+ \gamma \rightarrow e^+ \gamma^{*} \gamma \rightarrow e^+ W^- W^+$ and $e^+ e^-\rightarrow e^+ \gamma^{*} \gamma^{*} e^-
\rightarrow e^+ W^- W^+ e^-$ at the CLIC are obtained from:

\begin{eqnarray}
\sigma=\int f_{\gamma(\gamma^{*})}(x_1)f_{\gamma(\gamma^{*})}(x_2) d\hat{\sigma}_{\gamma(\gamma^{*}), \gamma(\gamma^{*})} dE_{1} dE_{2}.
\end{eqnarray}

For the computation of the total cross-section $\sigma(\Delta\kappa_\gamma, \lambda_\gamma, \sqrt{s})$ we have implemented the interactions
term in the CalcHEP package \cite{Belyaev}. The cross-section of photon fusion channel $\gamma \gamma \rightarrow W^- W^+$ [Figs. 4-5]
depends largely on the center-of-mass energies of the collider, as well as of the anomalous $\lambda_\gamma$ and $\Delta\kappa_\gamma$
couplings. With the center-of-mass energies $\sqrt{s}=0.380, 1.5, 3$ TeV and $-0.2\leq\lambda_\gamma\leq 0.2$, the photon fusion
cross-sections are: $\sigma(\gamma \gamma \rightarrow W^- W^+)=50$ pb, $5\times 10^3$ pb and $2\times 10^5$ pb, respectively.
For $-3\leq \Delta\kappa_\gamma\leq 3$, the cross-sections are $\sigma(\gamma \gamma \rightarrow W^- W^+)=4\times 10^2$ pb,
$2\times 10^4$ pb and $7\times 10^5$ pb. The production cross-sections $\sigma(e^+ \gamma \rightarrow e^+ \gamma^{*} \gamma
\rightarrow e^+ W^- W^+)=2$ pb, $2\times 10^2$ pb, $6\times 10^3$ pb for $-0.2\leq\lambda_\gamma\leq 0.2$ and $\sigma(e^+ \gamma
\rightarrow e^+ \gamma^{*} \gamma \rightarrow e^+ W^- W^+)=10$ pb, $3\times 10^2$ pb and $4\times 10^3$ pb for $-3\leq\Delta\kappa_\gamma\leq 3$,
and are presented in Figs. 6-7. For the process $e^+ e^-\rightarrow e^+ \gamma^{*} \gamma^{*} e^-\rightarrow e^+ W^- W^+ e^-$,
the production cross-section $\sigma(e^+ e^-\rightarrow e^+ \gamma^{*} \gamma^{*} e^-\rightarrow e^+ W^- W^+ e^-)=0.4$ pb, 5 pb,
$2\times 10^2$ pb for $\sqrt{s}=0.380, 1.5, 3$ TeV and $-0.2\leq\lambda_\gamma\leq 0.2$. Whereas that, for $-3\leq\Delta\kappa_\gamma\leq 3$
the cross-section is $\sigma(e^+ e^-\rightarrow e^+ \gamma^{*} \gamma^{*} e^-\rightarrow e^+ W^- W^+ e^-)=0.2$ pb, $10^2$ pb, $4\times 10^2$ pb
as are shown in Figs. 8 and 9. As mentioned above, the total cross-sections depends significantly on the center-of-mass energies
of the collider, as well as of the aTGC $\lambda_\gamma$ and $\leq\Delta\kappa_\gamma$.

As can be seen from Figs. 4-9, in the case of $\sqrt{s}=0.38$ TeV where the center-of mass energy is relatively low, there is
an asymmetry of the cross-section values relative to the negative and positive values of the aTGC $\lambda_\gamma$ and
$\Delta\kappa_\gamma$, due to the cross terms of the aTGC with the SM terms. This asymmetry decreased significantly
due to the reduction of the effect of the SM in increasing center-of-mass energies.

\subsection{Limits on the anomalous couplings $\Delta\kappa_\gamma$ and $\lambda_\gamma$ through the process $\gamma\gamma\rightarrow W^+ W^-$}

To illustrate the expected $95\%$ confidence intervals for the parameters $\Delta\kappa_\gamma$ and $\lambda_\gamma$, we adopted the $\chi^2$
method:

\begin{equation}
\chi^2(\Delta\kappa_\gamma, \lambda_\gamma )=\Biggl(\frac{\sigma_{SM}-\sigma_{BSM}(\sqrt{s}, \Delta\kappa_\gamma, \lambda_\gamma)}{\sigma_{SM}\sqrt{(\delta_{st})^2+(\delta_{sys})^2}}\Biggr)^2,
\end{equation}

\noindent where $\sigma_{BSM}(\sqrt{s}, \Delta\kappa_\gamma, \lambda_\gamma)$ and $\sigma_{SM}$ are the cross-section
with and the without anomalous couplings $\Delta\kappa_\gamma$ and $\lambda_\gamma$. $\delta_{st}=\frac{1}{\sqrt{N_{SM}}}$
is the statistical error and $\delta_{sys}$ is the systematic error. The number of events is given by
$N_{SM}={\cal L}_{int}\times \sigma_{SM}\times BR(W^{\pm}\to qq', l\nu_l)$, where ${\cal L}_{int}$ is the integrated
luminosity of the CLIC and $l=e^-, \mu^-$. For $W^+W^-$ pair production we classify their decay products according to
the decomposition of $W^\pm$. In this paper, we assume that one of the $W^\pm$ bosons decays leptonically and the other
hadronically for the signal. This phenomenon has already been studied by ATLAS and CMS Collaborations \cite{cmstop1,cmstop2,atlastop}.
Thus, we assume that the branching rations for $W^\pm$ decays are: $BR(W^\pm \to q q')=0.454$ for hadronic decays,
$BR(W^+ \to q q'; W^- \to l\nu_{e, \mu})=0.143$ for semi-leptonic decays and $BR(W^\pm \to l\nu_{e, \mu})=0.045$ for light leptonic decays.

We probe the potential of the CLIC to estimate the limits on the aTGC through the process $\gamma\gamma\rightarrow W^+ W^-$,
based $\gamma\gamma$ colliders with the benchmark parameters for each stage in the updated scenario. The observed $95\%$
confidence intervals for the aTGC $\Delta\kappa_\gamma$ and $\lambda_\gamma$ are shown in Tables III-V. The
confidence intervals for a given $\Delta\kappa_\gamma$ or $\lambda_\gamma$ parameter are computing while fixing the another
anomalous parameter to zero. The confidence intervals are shown separately for the leptonic, semi-leptonic and hadronic decays
channels of the $W^{\pm}$ bosons.

From Tables III-V, the best limits for $\Delta\kappa_\gamma$ and $\lambda_\gamma$, taken one coupling at a time, are given by:

\begin{eqnarray}
\Delta\kappa_\gamma &=& [-0.00023, 0.00023], \hspace{3mm}   \mbox{$95\%$ C.L.}, \nonumber\\
\lambda_\gamma &=& [-0.00034, 0.27038], \hspace{3mm}   \mbox{$95\%$ C.L.},
\end{eqnarray}

\begin{eqnarray}
\Delta\kappa_\gamma &=& [-0.00010, 0.00010], \hspace{3mm}   \mbox{$95\%$ C.L.}, \nonumber\\
\lambda_\gamma &=& [-0.00007, 0.00936], \hspace{3mm}   \mbox{$95\%$ C.L.},
\end{eqnarray}

\begin{eqnarray}
\Delta\kappa_\gamma &=& [-0.00007, 0.00007], \hspace{3mm}   \mbox{$95\%$ C.L.}, \nonumber\\
\lambda_\gamma &=& [-0.00004, 0.00102], \hspace{3mm}   \mbox{$95\%$ C.L.},
\end{eqnarray}

\noindent for the hadronic channel with $\sqrt{s}=0.380, 1.5, 3\hspace{0.8mm}{\rm TeV}$ and ${\cal L}= 1000, 2500, 5000\hspace{0.8mm}{\rm fb^{-1}}$,
respectively.

\begin{table}
\caption{: The expected $95\%$ confidence level for the anomalous couplings $\Delta\kappa_\gamma$ and $\lambda_\gamma$,
through the process $\gamma\gamma\rightarrow W^+ W^-$ for $\sqrt{s}=0.380\hspace{0.8mm}{\rm TeV}$. The leptonic, semi-leptonic
and hadronic channels of the $W^+W^-$ in the final state are considered. The confidence level for each parameter are
calculated while fixing the another parameter to zero.}
\begin{center}
\begin{tabular}{|c|c|c|c|c|}
\hline\hline
\multicolumn{5}{|c|}{$\sqrt{s}=$ 0.380 TeV, \hspace{1cm} $95\%$ C.L.} \\
\hline
&  &  \multicolumn{3}{c|} {Channel}  \\
\cline{3-5}
& ${\cal L} \, (\rm fb^{-1})$  &\hspace{1cm} Leptonic \hspace{1cm} &\hspace{1cm} Semi-leptonic \hspace{1cm} & \hspace{1cm} Hadronic\hspace{1cm} \\
\hline
\cline{1-5}
 & 100  &  [-0.00226, 0.00225] & [-0.00128, 0.00127] & [-0.00072, 0.00071]  \\
 & 300  &  [-0.00130, 0.00130] & [-0.00073, 0.00073] & [-0.00041, 0.00041]  \\
\hspace{0.5cm} $\Delta \kappa_\gamma$ \hspace{0.5cm}
 & 500  & [-0.00101, 0.00101]  &  [-0.00057, 0.00057] & [-0.00032, 0.00032] \\
 & 700  &  [-0.00085, 0.00085] &  [-0.00048, 0.00048] & [-0.00027, 0.00027] \\
 & 1000 &  [-0.00071, 0.00071] &  [-0.00040, 0.00040] & [-0.00023, 0.00023] \\
\hline
 & 100 & [-0.00342, 0.27331] & [-0.00193, 0.27190] & [-0.00109, 0.27109] \\
 & 300 & [-0.00198, 0.27195] & [-0.00111, 0.27112] & [-0.00063, 0.27065] \\
$\lambda_\gamma$
 & 500 & [-0.00154, 0.27152] & [-0.00086, 0.27088] & [-0.00048, 0.27052] \\
 & 700 & [-0.00130, 0.27130] & [-0.00073, 0.27075] & [-0.00041, 0.27045] \\
 & 1000 & [-0.00109, 0.27109] & [-0.00061, 0.27064] & [-0.00034, 0.27038] \\
\hline\hline
\end{tabular}
\end{center}
\end{table}

\begin{table}
\caption{ The expected $95\%$ confidence level for the anomalous couplings $\Delta\kappa_\gamma$ and $\lambda_\gamma$,
through the process $\gamma\gamma\rightarrow W^+ W^-$ for $\sqrt{s}=1.5\hspace{0.8mm}{\rm TeV}$. The leptonic, semi-leptonic
and hadronic channels of the $W^+W^-$ in the final state are considered. The confidence level for each parameter are
calculated while fixing the another parameter to zero.}
\begin{center}
\begin{tabular}{|c|c|c|c|c|}
\hline\hline
\multicolumn{5}{|c|}{$\sqrt{s}=$ 1.5 TeV, \hspace{1cm} $95\%$ C.L.} \\
\hline
&  &  \multicolumn{3}{c|} {Channel}  \\
\cline{3-5}
& ${\cal L} \, (\rm fb^{-1})$  &\hspace{1cm} Leptonic \hspace{1cm} &\hspace{1cm} Semi-leptonic \hspace{1cm} & \hspace{1cm} Hadronic\hspace{1cm} \\
\hline
\cline{1-5}
 & 100  &  [-0.00154, 0.00153] & [-0.00087, 0.00086] & [-0.00049, 0.00049]  \\
 & 500  &  [-0.00069, 0.00069] & [-0.00039, 0.00039] & [-0.00022, 0.00022]  \\
\hspace{0.5cm} $\Delta \kappa_\gamma$ \hspace{0.5cm}
 & 1000 & [-0.00048, 0.00048] &  [-0.00027, 0.00027] & [-0.00015, 0.00015] \\
 & 1500 & [-0.00040, 0.00040] &   [-0.00022, 0.00022] & [-0.00013, 0.00013] \\
 & 2500 & [-0.00031, 0.00031] &   [-0.00017, 0.00017] & [-0.00010, 0.00010] \\
\hline
 & 100 & [-0.00110, 0.01039] & [-0.00065, 0.00993] & [-0.00037, 0.00966] \\
 & 500 & [-0.00052, 0.00981] & [-0.00030, 0.00958] & [-0.00017, 0.00945] \\
$\lambda_\gamma$
 & 1000 & [-0.00037, 0.00966] & [-0.00021, 0.00950] & [-0.00012, 0.00940] \\
 & 1500 & [-0.00031, 0.00959] & [-0.00017, 0.00946] & [-0.00010, 0.00938] \\
 & 2500 & [-0.00024, 0.00952] & [-0.00013, 0.00942] & [-0.00007, 0.00936] \\
\hline\hline
\end{tabular}
\end{center}
\end{table}

\begin{table}
\caption{The expected $95\%$ confidence level for the anomalous couplings $\Delta\kappa_\gamma$ and $\lambda_\gamma$,
through the process $\gamma\gamma\rightarrow W^+ W^-$ for $\sqrt{s}=3\hspace{0.8mm}{\rm TeV}$. The leptonic, semi-leptonic
and hadronic channels of the $W^+W^-$ in the final state are considered. The confidence level for each parameter are
calculated while fixing the another parameter to zero.}
\begin{center}
\begin{tabular}{|c|c|c|c|c|}
\hline\hline
\multicolumn{5}{|c|}{$\sqrt{s}=$ 3 TeV, \hspace{1cm} $95\%$ C.L.} \\
\hline
&  &  \multicolumn{3}{c|} {Channel}  \\
\cline{3-5}
& ${\cal L} \, (\rm fb^{-1})$  &\hspace{1cm} Leptonic \hspace{1cm} &\hspace{1cm} Semi-leptonic \hspace{1cm} & \hspace{1cm} Hadronic\hspace{1cm} \\
\hline
\cline{1-5}
 & 100   &  [-0.00153, 0.00148] & [-0.00085, 0.00084] & [-0.00048, 0.00047]  \\
 & 1000  &  [-0.00048, 0.00047] & [-0.00027, 0.00027] & [-0.00015, 0.00015]  \\
\hspace{0.5cm} $\Delta \kappa_\gamma$ \hspace{0.5cm}
 & 3000  & [-0.00027, 0.00027] &  [-0.00015, 0.00015] & [-0.00009, 0.00009] \\
 & 4000  &  [-0.00024, 0.00024] &   [-0.00013, 0.00013] & [-0.00008, 0.00008] \\
 & 5000  &  [-0.00021, 0.00021] &   [-0.00012, 0.00012] & [-0.00007, 0.00007] \\
\hline
 & 100 & [-0.00058, 0.00156] & [-0.00037, 0.00136] & [-0.00024, 0.00122] \\
 & 1000 & [-0.00023, 0.00122] & [-0.00014, 0.00112] & [-0.00008, 0.00106] \\
$\lambda_\gamma$
 & 1000 & [-0.00014, 0.00113] & [-0.00008, 0.00107] & [-0.00005, 0.00104] \\
 & 4000 & [-0.00013, 0.00111] & [-0.00007, 0.00106] & [-0.00004, 0.00103] \\
 & 5000 & [-0.00011, 0.00110] & [-0.00006, 0.00105] & [-0.00004, 0.00102] \\
\hline\hline
\end{tabular}
\end{center}
\end{table}

\begin{table}
\caption{The expected $95\%$ confidence level for the anomalous couplings $\Delta\kappa_\gamma$ and $\lambda_\gamma$,
through the process $e^+ \gamma \rightarrow e^+ \gamma^{*} \gamma \rightarrow e^+ W^- W^+$ for $\sqrt{s}=0.380\hspace{0.8mm}{\rm TeV}$.
The leptonic, semi-leptonic and hadronic channels of the $W^+W^-$ in the final state are considered. The confidence level for each
parameter are calculated while fixing the another parameter to zero.}
\begin{center}
\begin{tabular}{|c|c|c|c|c|}
\hline\hline
\multicolumn{5}{|c|}{$\sqrt{s}=$ 0.380 TeV, \hspace{1cm} $95\%$ C.L.} \\
\hline
&  &  \multicolumn{3}{c|} {Channel}  \\
\cline{3-5}
& ${\cal L} \, (\rm fb^{-1})$  &\hspace{1cm} Leptonic \hspace{1cm} &\hspace{1cm} Semi-leptonic \hspace{1cm} & \hspace{1cm} Hadronic\hspace{1cm} \\
\hline
\cline{1-5}
 & 100  &  [-0.00121, 0.00118] & [-0.00678, 0.00667] & [-0.00377, 0.00376]  \\
 & 300  &  [-0.00697, 0.00685] & [-0.00390, 0.00387] & [-0.00219, 0.00218]  \\
\hspace{0.5cm} $\Delta \kappa_\gamma$ \hspace{0.5cm}
 & 500  &  [-0.00538, 0.00532] &  [-0.00302, 0.00300] & [-0.00169, 0.00169] \\
 & 700  &  [-0.00450, 0.00450] &  [-0.00253, 0.00253] & [-0.00143, 0.00143] \\
 & 1000 &  [-0.00377, 0.00377] &  [-0.00212, 0.00212] & [-0.00119, 0.00119] \\
\hline
 & 100 & [-0.01806, 0.31034] & [-0.01041, 0.30322] & [-0.00594, 0.29905] \\
 & 300 & [-0.01068, 0.30347] & [-0.00610, 0.29920] & [-0.00346, 0.29673] \\
$\lambda_\gamma$
 & 500 & [-0.00834, 0.30129] & [-0.00474, 0.29793] & [-0.00268, 0.29601] \\
 & 700 & [-0.00708, 0.30011] & [-0.00402, 0.29725] & [-0.00227, 0.29562] \\
 & 1000 & [-0.00594, 0.29905] & [-0.00337, 0.29665] & [-0.00190, 0.29527] \\
\hline\hline
\end{tabular}
\end{center}
\end{table}

\begin{table}
\caption{The expected $95\%$ confidence level for the anomalous couplings $\Delta\kappa_\gamma$ and $\lambda_\gamma$,
through the process $e^+ \gamma \rightarrow e^+ \gamma^{*} \gamma \rightarrow e^+ W^- W^+$ for $\sqrt{s}=1.5\hspace{0.8mm}{\rm TeV}$.
The leptonic, semi-leptonic and hadronic channels of the $W^+W^-$ in the final state are considered. The confidence level for each
parameter are calculated while fixing the another parameter to zero.}
\begin{center}
\begin{tabular}{|c|c|c|c|c|}
\hline\hline
\multicolumn{5}{|c|}{$\sqrt{s}=$ 1.5 TeV, \hspace{1cm} $95\%$ C.L.} \\
\hline
&  &  \multicolumn{3}{c|} {Channel}  \\
\cline{3-5}
& ${\cal L} \, (\rm fb^{-1})$  &\hspace{1cm} Leptonic \hspace{1cm} &\hspace{1cm} Semi-leptonic \hspace{1cm} & \hspace{1cm} Hadronic\hspace{1cm} \\
\hline
\cline{1-5}
 & 100  &  [-0.00439, 0.00438] & [-0.00249, 0.00247] & [-0.00140, 0.00139]  \\
 & 500  &  [-0.00198, 0.00197] & [-0.00111, 0.00111] & [-0.00062, 0.00062]  \\
\hspace{0.5cm} $\Delta \kappa_\gamma$ \hspace{0.5cm}
 & 1000 & [-0.00139, 0.00139] &  [-0.00078, 0.00078] & [-0.00044, 0.00044] \\
 & 1500  &  [-0.00114, 0.00114] &   [-0.00064, 0.00064] & [-0.00036, 0.00036] \\
 & 2500  &  [-0.00088, 0.00088] &   [-0.00050, 0.00050] & [-0.00028, 0.00028] \\
\hline
 & 100 & [-0.00397, 0.02482] & [-0.00238, 0.02330] & [-0.00140, 0.02235] \\
 & 500 & [-0.00193, 0.02286] & [-0.00112, 0.02208] & [-0.00064, 0.02162] \\
$\lambda_\gamma$
 & 1000 & [-0.00140, 0.02235] & [-0.00081, 0.02178] & [-0.00046, 0.02140] \\
 & 1500 & [-0.00115, 0.02211] & [-0.00066, 0.02164] & [-0.00037, 0.02136] \\
 & 2500 & [-0.00090, 0.02187] & [-0.00051, 0.02150] & [-0.00029, 0.02128] \\
\hline\hline
\end{tabular}
\end{center}
\end{table}

\begin{table}
\caption{The expected $95\%$ confidence level for the anomalous couplings $\Delta\kappa_\gamma$ and $\lambda_\gamma$,
through the process $e^+ \gamma \rightarrow e^+ \gamma^{*} \gamma \rightarrow e^+ W^- W^+$ for $\sqrt{s}=3\hspace{0.8mm}{\rm TeV}$.
The leptonic, semi-leptonic and hadronic channels of the $W^+W^-$ in the final state are considered. The confidence level for
each parameter are calculated while fixing the another parameter to zero.}
\begin{center}
\begin{tabular}{|c|c|c|c|c|}
\hline\hline
\multicolumn{5}{|c|}{$\sqrt{s}=$ 3 TeV, \hspace{1cm} $95\%$ C.L.} \\
\hline
&  &  \multicolumn{3}{c|} {Channel}  \\
\cline{3-5}
& ${\cal L} \, (\rm fb^{-1})$  &\hspace{1cm} Leptonic \hspace{1cm} &\hspace{1cm} Semi-leptonic \hspace{1cm} & \hspace{1cm} Hadronic\hspace{1cm} \\
\hline
\cline{1-5}
 & 100   &  [-0.00352, 0.00341] & [-0.00196, 0.00193] & [-0.00110, 0.00109]  \\
 & 1000  &  [-0.00110, 0.00109] & [-0.00062, 0.00061] & [-0.00035, 0.00035]  \\
\hspace{0.5cm} $\Delta \kappa_\gamma$ \hspace{0.5cm}
 & 3000  &  [-0.00063, 0.00063] & [-0.00035, 0.00035] & [-0.00020, 0.00020] \\
 & 4000  &  [-0.00055, 0.00055] & [-0.00031, 0.00031] & [-0.00017, 0.00017] \\
 & 5000  &  [-0.00049, 0.00049] & [-0.00027, 0.00027] & [-0.00015, 0.00015] \\
\hline
 & 100  & [-0.00195, 0.00522] & [-0.00126, 0.00453] & [-0.00079, 0.00406] \\
 & 1000 & [-0.00079, 0.00406] & [-0.00048, 0.00375] & [-0.00028, 0.00355] \\
$\lambda_\gamma$
 & 3000 & [-0.00049, 0.00376] & [-0.00029, 0.00356] & [-0.00017, 0.00344] \\
 & 4000 & [-0.00043, 0.00370] & [-0.00025, 0.00352] & [-0.00014, 0.00342] \\
 & 5000 & [-0.00039, 0.00366] & [-0.00023, 0.00350] & [-0.00013, 0.00340] \\
\hline\hline
\end{tabular}
\end{center}
\end{table}

\begin{table}
\caption{The expected $95\%$ confidence level for the anomalous couplings $\Delta\kappa_\gamma$ and $\lambda_\gamma$,
through the process $e^+ e^-\rightarrow e^+ \gamma^{*} \gamma^{*} e^-\rightarrow e^+ W^- W^+ e^-$ for $\sqrt{s}=0.380\hspace{0.8mm}{\rm TeV}$.
The leptonic, semi-leptonic and hadronic channels of the $W^+W^-$ in the final state are considered. The confidence level for each parameter
are calculated while fixing the another parameter to zero.}
\begin{center}
\begin{tabular}{|c|c|c|c|c|}
\hline\hline
\multicolumn{5}{|c|}{$\sqrt{s}=$ 0.380 TeV, \hspace{1cm} $95\%$ C.L.} \\
\hline
&  &  \multicolumn{3}{c|} {Channel}  \\
\cline{3-5}
& ${\cal L} \, (\rm fb^{-1})$  &\hspace{1cm} Leptonic \hspace{1cm} &\hspace{1cm} Semi-leptonic \hspace{1cm} & \hspace{1cm} Hadronic\hspace{1cm} \\
\hline
\cline{1-5}
 & 100  &  [-0.07089, 0.06104] & [-0.03840, 0.03532] & [-0.02117, 0.02020]  \\
 & 300  &  [-0.03949, 0.03624] & [-0.02175, 0.02073] & [-0.01209, 0.01177]  \\
\hspace{0.5cm} $\Delta \kappa_\gamma$ \hspace{0.5cm}
 & 500  &  [-0.03027, 0.02832] &  [-0.01675, 0.01614] & [-0.00934, 0.00915] \\
 & 700  &  [-0.02544, 0.02405] &  [-0.01412, 0.01368] & [-0.00788, 0.00774] \\
 & 1000 &  [-0.02118, 0.02021] &  [-0.01178, 0.01148] & [-0.00658, 0.00649] \\
\hline
 & 100 & [-0.08393, 0.38546] & [-0.05161, 0.35633] & [-0.03084, 0.33743] \\
 & 300 & [-0.05282, 0.35742] & [-0.03160, 0.33813] & [-0.01848, 0.32613] \\
$\lambda_\gamma$
 & 500 & [-0.04219, 0.34778] & [-0.02497, 0.33207] & [-0.01449, 0.32247] \\
 & 700 & [-0.03629, 0.34240] & [-0.02134, 0.32875] & [-0.01233, 0.32048] \\
 & 1000 & [-0.03085, 0.33745] & [-0.01803, 0.32572] & [-0.01038, 0.31869] \\
\hline\hline
\end{tabular}
\end{center}
\end{table}

\begin{table}
\caption{The expected $95\%$ confidence level for the anomalous couplings $\Delta\kappa_\gamma$ and $\lambda_\gamma$,
through the process $e^+ e^-\rightarrow e^+ \gamma^{*} \gamma^{*} e^-\rightarrow e^+ W^- W^+ e^-$ for $\sqrt{s}=1.5\hspace{0.8mm}{\rm TeV}$.
The leptonic, semi-leptonic and hadronic channels of the $W^+W^-$ in the final state are considered. The confidence level for each
parameter are calculated while fixing the another parameter to zero.}
\begin{center}
\begin{tabular}{|c|c|c|c|c|}
\hline\hline
\multicolumn{5}{|c|}{$\sqrt{s}=$ 1.5 TeV, \hspace{1cm} $95\%$ C.L.} \\
\hline
&  &  \multicolumn{3}{c|} {Channel}  \\
\cline{3-5}
& ${\cal L} \, (\rm fb^{-1})$  &\hspace{1cm} Leptonic \hspace{1cm} &\hspace{1cm} Semi-leptonic \hspace{1cm} & \hspace{1cm} Hadronic\hspace{1cm} \\
\hline
\cline{1-5}
 & 100  &  [-0.01668, 0.01578] & [-0.00926, 0.00898] & [-0.00516, 0.00508]  \\
 & 500  &  [-0.00734, 0.00716] & [-0.00410, 0.00405] & [-0.00230, 0.00228]  \\
\hspace{0.5cm} $\Delta \kappa_\gamma$ \hspace{0.5cm}
 & 1000 & [-0.00517, 0.00508] &  [-0.00286, 0.00287] & [-0.00162, 0.00162] \\
 & 1500 &  [-0.00421, 0.00416] &   [-0.00236, 0.00234] & [-0.00132, 0.00132] \\
 & 2500 &  [-0.00326, 0.00322] &   [-0.00182, 0.00182] & [-0.00102, 0.00102] \\
\hline
 & 100 & [-0.01395, 0.04782] & [-0.00878, 0.04316] & [-0.00536, 0.04002] \\
 & 500 & [-0.00725, 0.04175] & [-0.00437, 0.03911] & [-0.00257, 0.03744] \\
$\lambda_\gamma$
 & 1000 & [-0.00536, 0.04000] & [-0.00319, 0.03801] & [-0.00185, 0.03678] \\
 & 1500 & [-0.00448, 0.03921] & [-0.00264, 0.03750] & [-0.00153, 0.03647] \\
 & 2500 & [-0.00355, 0.03835] & [-0.00207, 0.03698] & [-0.00119, 0.03616] \\
\hline\hline
\end{tabular}
\end{center}
\end{table}

\begin{table}
\caption{The expected $95\%$ confidence level for the anomalous couplings $\Delta\kappa_\gamma$ and $\lambda_\gamma$,
through the process $e^+ e^-\rightarrow e^+ \gamma^{*} \gamma^{*} e^-\rightarrow e^+ W^- W^+ e^-$ for $\sqrt{s}=3\hspace{0.8mm}{\rm TeV}$.
The leptonic, semi-leptonic and hadronic channels of the $W^+W^-$ in the final state are considered. The confidence level for each
parameter are calculated while fixing the another parameter to zero.}
\begin{center}
\begin{tabular}{|c|c|c|c|c|}
\hline\hline
\multicolumn{5}{|c|}{$\sqrt{s}=$ 3 TeV, \hspace{1cm} $95\%$ C.L.} \\
\hline
&  &  \multicolumn{3}{c|} {Channel}  \\
\cline{3-5}
& ${\cal L} \, (\rm fb^{-1})$  &\hspace{1cm} Leptonic \hspace{1cm} &\hspace{1cm} Semi-leptonic \hspace{1cm} & \hspace{1cm} Hadronic\hspace{1cm} \\
\hline
\cline{1-5}
 & 100   &  [-0.01122, 0.01061] & [-0.00622, 0.00603] & [-0.00347, 0.00341]  \\
 & 1000  &  [-0.00347, 0.00342] & [-0.00194, 0.00193] & [-0.00109, 0.00109]  \\
\hspace{0.5cm} $\Delta \kappa_\gamma$ \hspace{0.5cm}
 & 3000  &  [-0.00199, 0.00198] & [-0.00112, 0.00112] & [-0.00062, 0.00063] \\
 & 4000  &  [-0.00173, 0.00172] & [-0.00097, 0.00097] & [-0.00054, 0.00054] \\
 & 5000  &  [-0.00154, 0.00154] & [-0.00086, 0.00086] & [-0.00048, 0.00049] \\
\hline
 & 100  & [-0.00621, 0.01341] & [-0.00414, 0.01140] & [-0.00267, 0.00997] \\
 & 1000 & [-0.00268, 0.00997] & [-0.00167, 0.00899] & [-0.00101, 0.00835] \\
$\lambda_\gamma$
 & 3000 & [-0.00171, 0.00903] & [-0.00104, 0.00837] & [-0.00061, 0.00795] \\
 & 4000 & [-0.00151, 0.00884] & [-0.00091, 0.00825] & [-0.00053, 0.00788] \\
 & 5000 & [-0.00137, 0.00870] & [-0.00082, 0.00816] & [-0.00048, 0.00782] \\
\hline\hline
\end{tabular}
\end{center}
\end{table}

For the other luminosity stages of the CLIC, as well as of the other decay channels of the $W^\pm$ bosons, the limits for
$\Delta\kappa_\gamma$ and $\lambda_\gamma$ are weaker than those corresponding to Eqs. (18)-(20), however there are also
competitive with the experimental limits which are shown in Table I for the ATLAS, CMS, CDF, D0, ALEP, DELPHI, L3 and OPAL
Collaborations, as well as with the corresponding phenomenological limits obtained for the future ILC and the CEPC. It is
worth mentioning that for all the CLIC energy stages, the process $\gamma\gamma\rightarrow W^+ W^-$ gives strong limits to
the anomalous couplings $\Delta\kappa_\gamma$ and $\lambda_\gamma$, as shown in Eqs. (18)-(20) and Tables III-V. These results
show the strong benefit of several energy stages for the CLIC physics potential. In addition, the operation at high energy
significantly improves the limits to the anomalous couplings.

\subsection{Limits on the anomalous couplings $\Delta\kappa_\gamma$ and $\lambda_\gamma$ through the process
$e^+ \gamma \rightarrow e^+ \gamma^{*} \gamma \rightarrow e^+ W^- W^+$}

The anomalous couplings $\Delta\kappa_\gamma$ and $\lambda_\gamma$ to the photon and $W^\pm$ bosons are precisely predicted
by the SM but may receive substantial corrections from BSM physics. Here we have shown the physics potential of $W^+W^-$ pair
production through the process $e^+ \gamma \rightarrow e^+ \gamma^{*} \gamma \rightarrow e^+ W^- W^+$ in $\gamma\gamma^*$
collisions and using effective Lagrangian. The effective Lagrangian formalism extends the SM Lagrangian to include interaction
operators of higher dimension. The leading effects are captured by dimension-six operators weighted by the coefficients $C_i/\Lambda^2$
for dimensionless couplings $C_i$, as shown in Eq. (1). With this focus and with the clean experimental environment of the CLIC the
measurements on the anomalous couplings $\Delta\kappa_\gamma$ and $\lambda_\gamma$ to be better exploited.

Already the first CLIC stage provides an important set of measurements on the anomalous couplings $\Delta\kappa_\gamma$ and
$\lambda_\gamma$ using the $e^+ \gamma \rightarrow e^+ \gamma^{*} \gamma \rightarrow e^+ W^- W^+$ process, as shown in Table VI.
Next, we present the most significant limits for the anomalous couplings coming from the vertex $W^+W^-\gamma$, for all energy stages
of the CLIC (see Tables VI-VIII):

\begin{eqnarray}
\Delta\kappa_\gamma &=& [-0.00119, 0.00119], \hspace{3mm}   \mbox{$95\%$ C.L.}, \nonumber\\
\lambda_\gamma &=& [-0.00190, 0.29527], \hspace{3mm}   \mbox{$95\%$ C.L.},
\end{eqnarray}

\begin{eqnarray}
\Delta\kappa_\gamma &=& [-0.00028, 0.00028], \hspace{3mm}   \mbox{$95\%$ C.L.}, \nonumber\\
\lambda_\gamma &=& [-0.00029, 0.02128], \hspace{3mm}   \mbox{$95\%$ C.L.},
\end{eqnarray}

\begin{eqnarray}
\Delta\kappa_\gamma &=& [-0.00015, 0.00015], \hspace{3mm}   \mbox{$95\%$ C.L.}, \nonumber\\
\lambda_\gamma &=& [-0.00013, 0.00340], \hspace{3mm}   \mbox{$95\%$ C.L.},
\end{eqnarray}

\noindent for the hadronic channel with $\sqrt{s}=0.380, 1.5, 3\hspace{0.8mm}{\rm TeV}$ and ${\cal L}= 1000, 2500, 5000\hspace{0.8mm}{\rm fb^{-1}}$,
respectively. The limits on $\Delta\kappa_\gamma$ and $\lambda_\gamma$ given in Eqs. (21)-(23) are competitive with those shown in Table I, and in
some cases our limits are stronger.

\subsection{Limits on the anomalous couplings $\Delta\kappa_\gamma$ and $\lambda_\gamma$ through the process
$e^+ e^-\rightarrow e^+ \gamma^{*} \gamma^{*} e^-\rightarrow e^+ W^- W^+ e^-$}

The WWA of quasi-real quanta, is best known for its application to radiation during elementary particle collisions. As we mentioned earlier,
other well-known applications of the linear colliders are the processes $e\gamma^{*}$, $\gamma \gamma^{*}$ and $\gamma^{*} \gamma^{*}$.

Combining the WWA with the characteristics of the future CLIC, such as high energies, high luminosities, and clean experimental environments,
we probe limits on the anomalous couplings $\Delta\kappa_\gamma$ and $\lambda_\gamma$ of the $W^\pm$ bosons. The limits obtained for
$\Delta\kappa_\gamma$ and $\lambda_\gamma$ through the process $e^+ e^-\rightarrow e^+ \gamma^{*} \gamma^{*} e^-\rightarrow e^+ W^- W^+ e^-$
are shown in Tables IX-XI. From these tables, the most notable limits are the following:

\begin{eqnarray}
\Delta\kappa_\gamma &=& [-0.00658, 0.00649], \hspace{3mm}   \mbox{$95\%$ C.L.}, \nonumber\\
\lambda_\gamma &=& [-0.01038, 0.31869], \hspace{3mm}   \mbox{$95\%$ C.L.},
\end{eqnarray}

\begin{eqnarray}
\Delta\kappa_\gamma &=& [-0.00102, 0.00102], \hspace{3mm}   \mbox{$95\%$ C.L.}, \nonumber\\
\lambda_\gamma &=& [-0.00119, 0.03616], \hspace{3mm}   \mbox{$95\%$ C.L.},
\end{eqnarray}

\begin{eqnarray}
\Delta\kappa_\gamma &=& [-0.00048, 0.00049], \hspace{3mm}   \mbox{$95\%$ C.L.}, \nonumber\\
\lambda_\gamma &=& [-0.00048, 0.00782], \hspace{3mm}   \mbox{$95\%$ C.L.},
\end{eqnarray}

\noindent for the hadronic channel with $\sqrt{s}=0.380, 1.5, 3\hspace{0.8mm}{\rm TeV}$ and ${\cal L}= 1000, 2500, 5000\hspace{0.8mm}{\rm fb^{-1}}$,
respectively.

To conclude these subsections, it is worth mentioning that already after the initial energy stage, in many cases (leptonic,
semi-leptonic channels and hadronic, as well as for different luminosities) the CLIC precision is significantly better than
for the results shown in Table I, and improves further with higher energy running.

\section{Summary of the achievable precision on the anomalous couplings $\Delta\kappa_\gamma$ and $\lambda_\gamma$}

To complement our study on the anomalous couplings $\Delta\kappa_\gamma$ and $\lambda_\gamma$ through the processes
$\gamma\gamma\rightarrow W^+ W^-$, $e^+ \gamma \rightarrow e^+ \gamma^{*} \gamma \rightarrow e^+ W^- W^+$ and
$e^+ e^-\rightarrow e^+ \gamma^{*} \gamma^{*} e^-\rightarrow e^+ W^- W^+ e^-$, for the three CLIC energy stages (see Table II),
we present the bar graphs given by Figs. 10-17.

It is worth mentioning that, the sources of systematic uncertainty in our measurements can be due to uncertainties in the
integrated luminosity ${\cal L}$, in factors that corrects for experimental acceptance and efficiencies, different background
sources, particle identification and misstagging. To reduce the systematic uncertainties in our study, we used the systematic 
uncertainties $\delta_{sys}= 0\%, 3\%, 5\%$ as a benchmark in our computation.

Comparison of precisions at the CLIC to the anomalous couplings $\lambda_\gamma$, $\Delta\kappa_\gamma$ for center-of-mass energy $\sqrt{s}=0.380\hspace{0.8mm}{\rm TeV}$ and luminosities ${\cal L}=100, 500, 1000\hspace{0.8mm}{\rm fb^{-1}}$ are shown in Figs. 10 and 11.
The figures covers the three processes $\gamma\gamma\rightarrow W^+ W^-$, $e^+ \gamma \rightarrow e^+ \gamma^{*} \gamma \rightarrow e^+ W^- W^+$
and $e^+ e^-\rightarrow e^+ \gamma^{*} \gamma^{*} e^-\rightarrow e^+ W^- W^+ e^-$. Figs. 10 and 11 show that the initial stage of CLIC is already
very complementary to the experimental limits at $95\%$ C.L. on the aTGC $\Delta\kappa_\gamma$ and $\lambda_\gamma$ from the present
and future colliders, as shown in Table I.

The high energy stages, which are unique to the CLIC among all proposed $e^+e^-$ colliders, are found to be crucial for the precision
measurements. This is illustrated in Figs. 12-15, for the comparison of precisions at the CLIC to the anomalous coupling $\lambda_\gamma$
for center-of-mass energies $\sqrt{s}=1.5, 3\hspace{0.8mm}{\rm TeV}$ and luminosities ${\cal L}=100, 1000, 2500, 3000, 5000\hspace{0.8mm}{\rm fb^{-1}}$.
The figures cover the three processes $\gamma\gamma\rightarrow W^+ W^-$, $e^+ \gamma \rightarrow e^+ \gamma^{*} \gamma \rightarrow e^+ W^- W^+$
and $e^+ e^-\rightarrow e^+ \gamma^{*} \gamma^{*} e^-\rightarrow e^+ W^- W^+ e^-$.

Figs. 16 and 17, show the comparison of precisions at the CLIC for the anomalous couplings $\lambda_\gamma$ and $\Delta\kappa_\gamma$
for $\sqrt{s}=3\hspace{0.8mm}{\rm TeV}$ and ${\cal L}=5000\hspace{0.8mm}{\rm fb^{-1}}$,
under three systematic uncertainty scenarios, $\delta_{sys}=0, 3, 5\%$. The figures cover the three processes
$\gamma\gamma\rightarrow W^+ W^-$, $e^+ \gamma \rightarrow e^+ \gamma^{*} \gamma \rightarrow e^+ W^- W^+$ and
$e^+ e^-\rightarrow e^+ \gamma^{*} \gamma^{*} e^-\rightarrow e^+ W^- W^+ e^-$, with pure leptonic decays.
Black markers correspond to precision of $\lambda_\gamma$ ($\Delta\kappa_\gamma$) in $\gamma\gamma$ collisions,
green markers correspond to results from $\gamma\gamma^*$ collisions, and blue markers give the results for the case of
$\gamma^*\gamma^*$ collisions.

\section{Conclusions}

\vspace{3mm}

CLIC has the sensitivity to a large set of the anomalous couplings, in particular for the aTGC $W^+W^-\gamma$. In addition to the
favorable experimental conditions of CLIC such as clean environments, low background, high-energy and high-luminosity, as well
as operate in $\gamma\gamma$, $\gamma\gamma^*$ and $\gamma^*\gamma^*$ collision modes. All these features of the CLIC allow to
explorer the limits on the anomalous couplings $\Delta\kappa_\gamma$ and $\lambda_\gamma$ through the processes
$\gamma\gamma\rightarrow W^+ W^-$, $e^+ \gamma \rightarrow e^+ \gamma^{*} \gamma \rightarrow e^+ W^- W^+$
and $e^+ e^-\rightarrow e^+ \gamma^{*} \gamma^{*} e^-\rightarrow e^+ W^- W^+ e^-$ for the three CLIC energy stages.

The CLIC capability to perform multiple competitive probings of the anomalous couplings $\Delta\kappa_\gamma$ and $\lambda_\gamma$
allows robust conclusions to be drawn (see Sections III and IV). In this regard, our results are summarized through a set of
Tables III-XI and Figs. 10-17. From these tables and figures, the indicative CLIC reach for new physics, is given for $\Delta\kappa_\gamma$
and $\lambda_\gamma$ for the full CLIC physics program covering the three center-of-mass energy stages. The best limits are at $95\%$ C.L.:
$\Delta\kappa_\gamma = [-7, 7]\times 10^{-5}$, $\lambda_\gamma = [-0.4, 10.2]\times 10^{-4}$, through the signal
$\gamma\gamma\rightarrow W^+ W^-$; $\Delta\kappa_\gamma = [-1.5, 1.5]\times 10^{-4}$, $\lambda_\gamma = [-0.13, 3.40]\times 10^{-3}$
for the process $e^+ \gamma \rightarrow e^+ \gamma^{*} \gamma \rightarrow e^+ W^- W^+$, and $\Delta\kappa_\gamma = [-4.8, 4.9]\times 10^{-4}$,
$\lambda_\gamma = [-0.48, 7.82]\times 10^{-3}$ for the mode $e^+ e^-\rightarrow e^+ \gamma^{*} \gamma^{*} e^-\rightarrow e^+ W^- W^+ e^-$,
respectively. All these results are for the hadronic channels of the $W^\pm$ bosons.

In conclusion, our limits on the anomalous couplings $\Delta\kappa_\gamma$ and $\lambda_\gamma$ indicate that the CLIC for
its three energy stages can measure these couplings to a level of precision that exceeds that of the ATLAS, CMS, CDF, D0,
ALEPH, DELPHI, L3 and OPAL Collaborations by more than ${\cal O}(10^{-3}-10^{-2})$ order of magnitude. In this context,
the innovative project such as the CLIC with BSM physics searches  is highly desirable, and guaranteed outcome of precision
measurements.

\vspace{1.5cm}

\begin{center}
{\bf Acknowledgements}
\end{center}

A. G. R. and M. A. H. R. acknowledge support from SNI and PROFOCIE (M\'exico).

\vspace{2cm}


\newpage

\begin{figure}[H]
\centerline{\scalebox{0.7}{\includegraphics{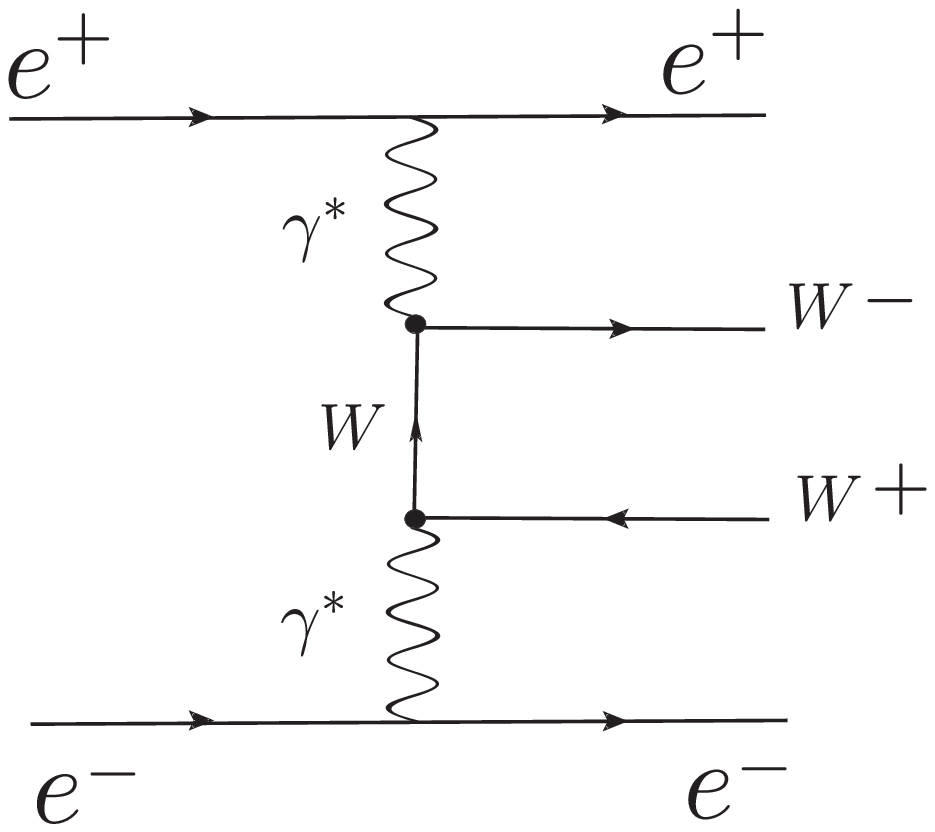}}}
\caption{ \label{fig:gamma} A schematic diagram for the process $e^+e^- \to e^+ \gamma^*\gamma^* e^- \rightarrow e^+ W^- W^+e^-$
via the subprocess $\gamma^* \gamma^* \rightarrow W^+ W^-$.}
\end{figure}

\begin{figure}[H]
\centerline{\scalebox{0.7}{\includegraphics{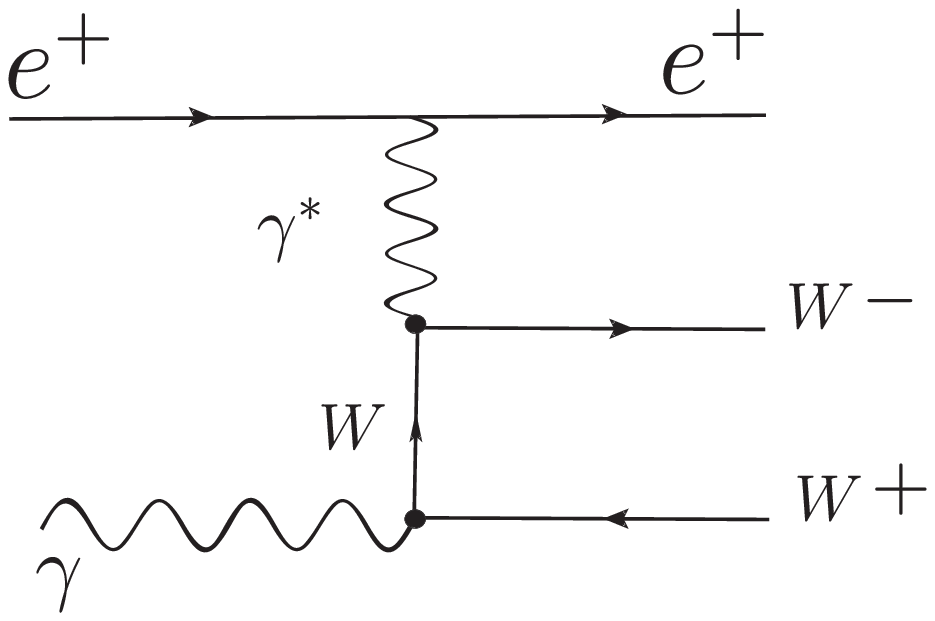}}}
\caption{ \label{fig:gamma} A schematic diagram for the process $e^+ \gamma \rightarrow e^+ \gamma^{*} \gamma \rightarrow e^+ W^+W^-$
via the subprocess $\gamma \gamma^{*} \rightarrow W^+ W^-$.}
\end{figure}

\begin{figure}[H]
\centerline{\scalebox{0.7}{\includegraphics{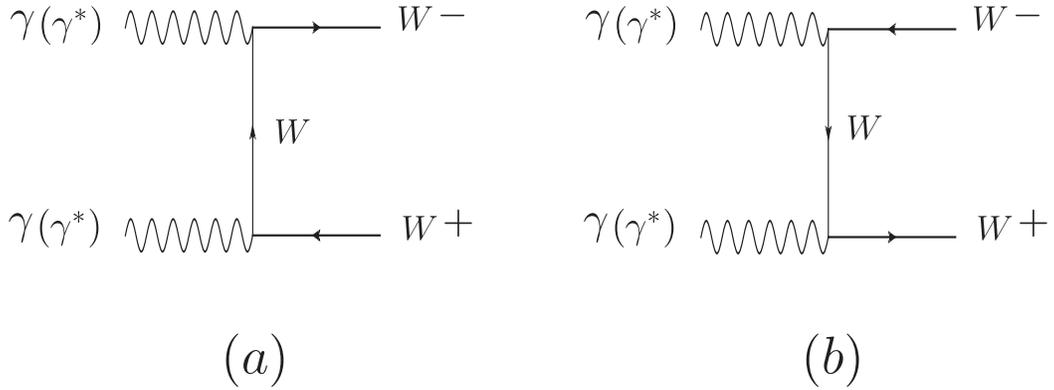}}}
\caption{ \label{fig:gamma1} Feynman diagrams contributing to the process
$\gamma\gamma \to W^+ W^-$ and the subprocesses $\gamma\gamma^* \to W^+ W^-$ and $\gamma^*\gamma^* \to W^+W^-$.}
\end{figure}

\begin{figure}[H]
\centerline{\scalebox{1.5}{\includegraphics{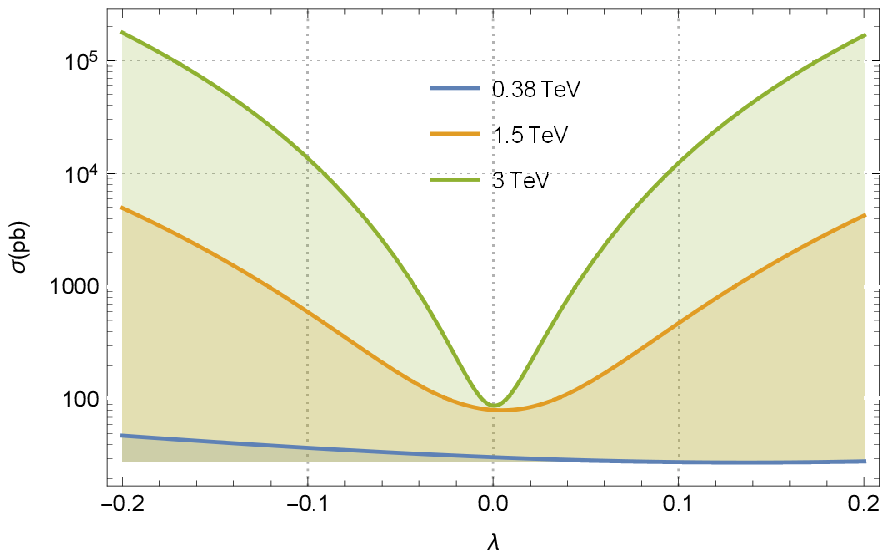}}}
\caption{The total cross-sections of the process $\gamma\gamma\rightarrow W^+ W^-$ as a function of $\lambda_\gamma$
for center-of-mass energies of $\sqrt{s}=0.380, 1.5, 3\hspace{0.8mm}$ TeV at the CLIC.}
\label{Fig.3}
\end{figure}

\begin{figure}[H]
\centerline{\scalebox{1.5}{\includegraphics{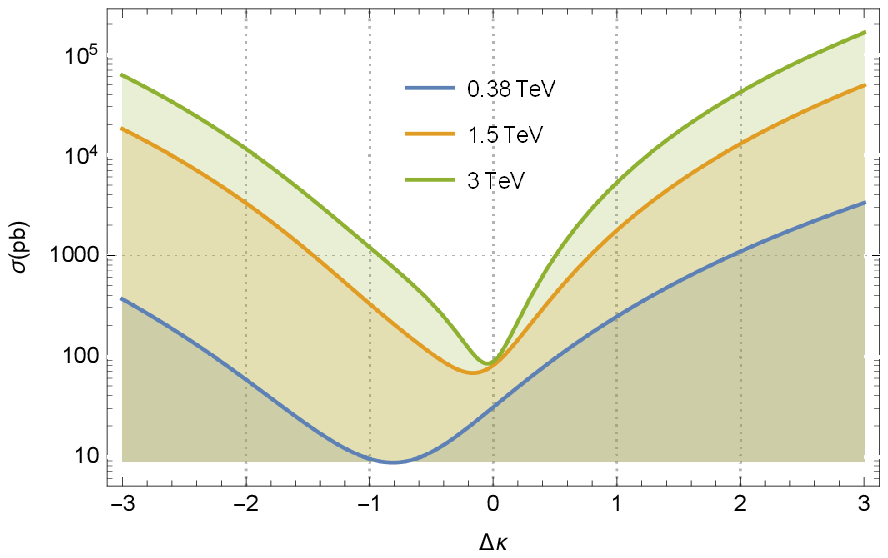}}}
\caption{The total cross-sections of the process $\gamma\gamma\rightarrow W^+ W^-$ as a function of $\Delta\kappa_\gamma$
for center-of-mass energies of $\sqrt{s}=0.380, 1.5, 3\hspace{0.8mm}$ TeV at the CLIC.}
\label{Fig.3}
\end{figure}

\begin{figure}[H]
\centerline{\scalebox{1.5}{\includegraphics{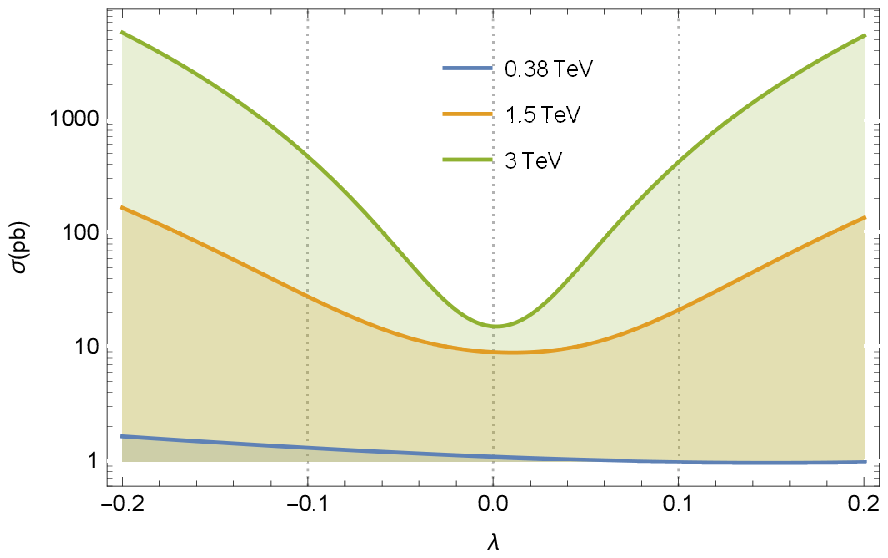}}}
\caption{The same as in Fig. 4, but for the process
$e^+ \gamma \rightarrow e^+ \gamma^{*} \gamma \rightarrow e^+ W^- W^+$.}
\label{Fig.3}
\end{figure}

\begin{figure}[H]
\centerline{\scalebox{1.5}{\includegraphics{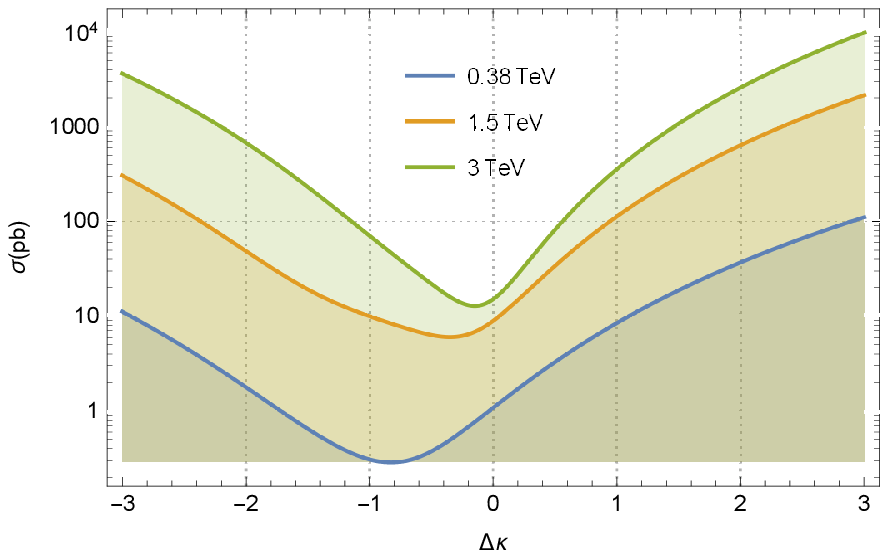}}}
\caption{The same as in Fig. 5, but for the process
$e^+ \gamma \rightarrow e^+ \gamma^{*} \gamma \rightarrow e^+ W^- W^+$.}
\label{Fig.3}
\end{figure}

\begin{figure}[H]
\centerline{\scalebox{1.5}{\includegraphics{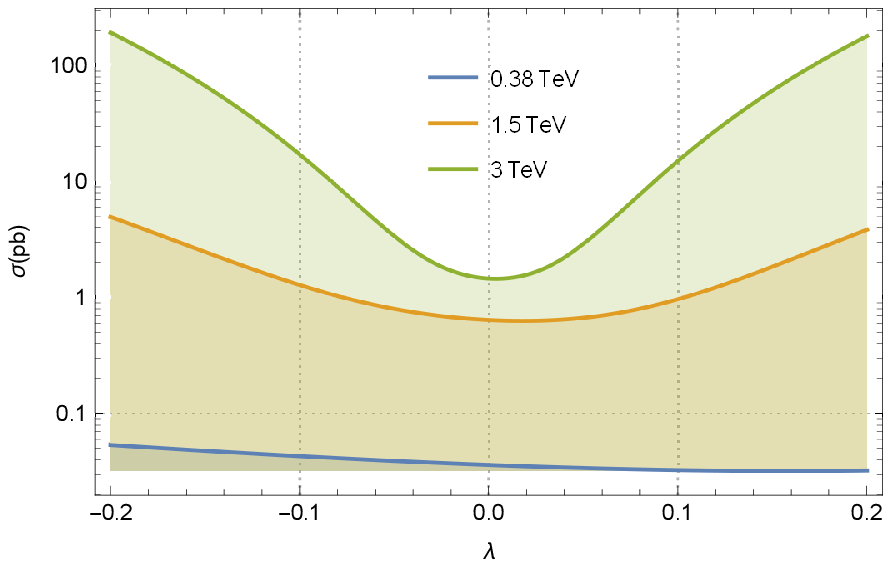}}}
\caption{The same as in Fig. 4, but for the process
$e^+ e^-\rightarrow e^+ \gamma^{*} \gamma^{*} e^-\rightarrow e^+ W^- W^+ e^-$.}
\label{Fig.3}
\end{figure}

\begin{figure}[H]
\centerline{\scalebox{1.5}{\includegraphics{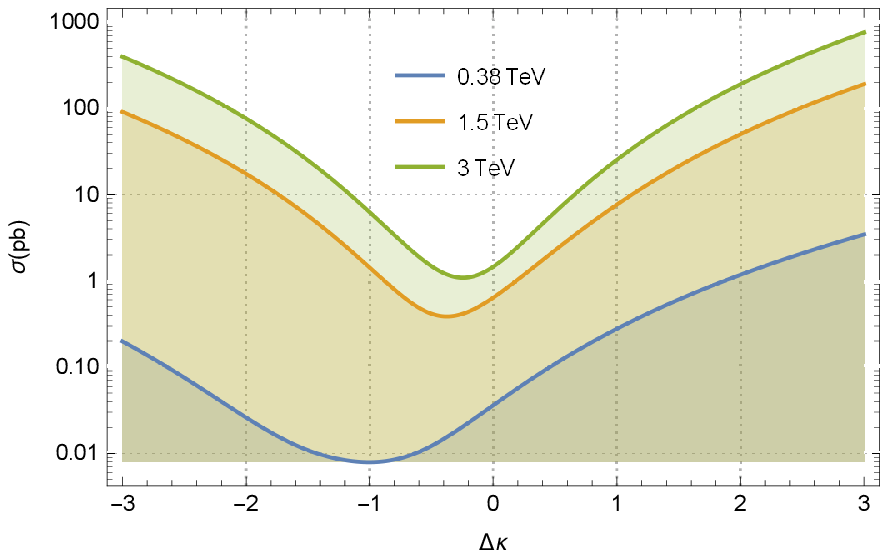}}}
\caption{The same as in Fig. 5, but for the process
$e^+ e^-\rightarrow e^+ \gamma^{*} \gamma^{*} e^-\rightarrow e^+ W^- W^+ e^-$.}
\label{Fig.3}
\end{figure}

\begin{figure}[H]
\centerline{\scalebox{1}{\includegraphics{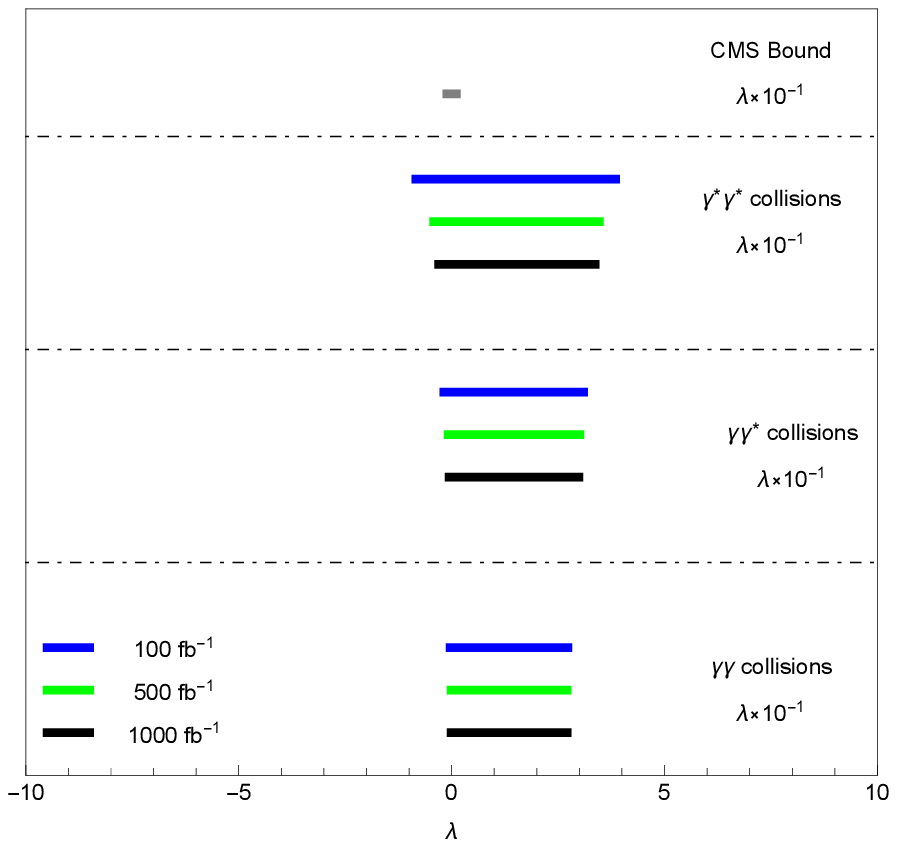}}}
\caption{ \label{fig:gamma2} Comparison of precisions at the CLIC to the anomalous coupling  $\lambda_\gamma$
for center-of-mass energy $\sqrt{s}=0.380\hspace{0.8mm}{\rm TeV}$ and luminosities
${\cal L}=100, 500, 1000\hspace{0.8mm}{\rm fb^{-1}}$. The figure covers the three processes
$\gamma\gamma\rightarrow W^+ W^-$, $e^+ \gamma \rightarrow e^+ \gamma^{*} \gamma \rightarrow e^+ W^- W^+$
and $e^+ e^-\rightarrow e^+ \gamma^{*} \gamma^{*} e^-\rightarrow e^+ W^- W^+ e^-$. We include the CMS bound.}
\end{figure}

\begin{figure}[H]
\centerline{\scalebox{1}{\includegraphics{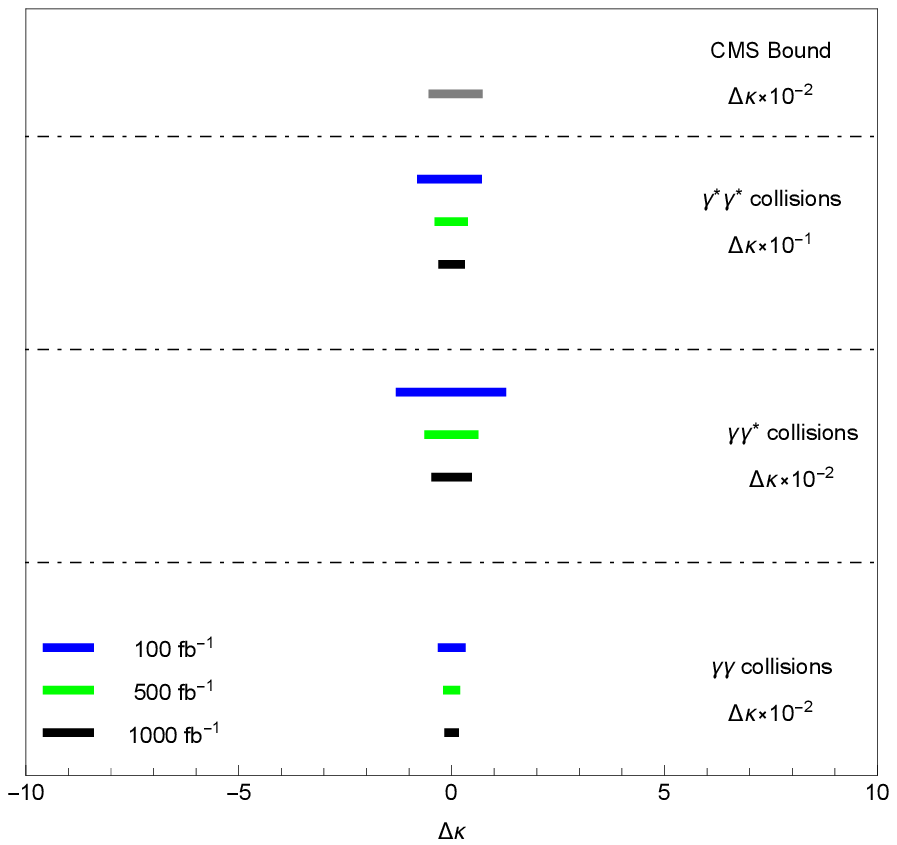}}}
\caption{ \label{fig:gamma3} The same as in Fig. 10, but for $\Delta\kappa_\gamma$.}
\end{figure}

\begin{figure}[H]
\centerline{\scalebox{1}{\includegraphics{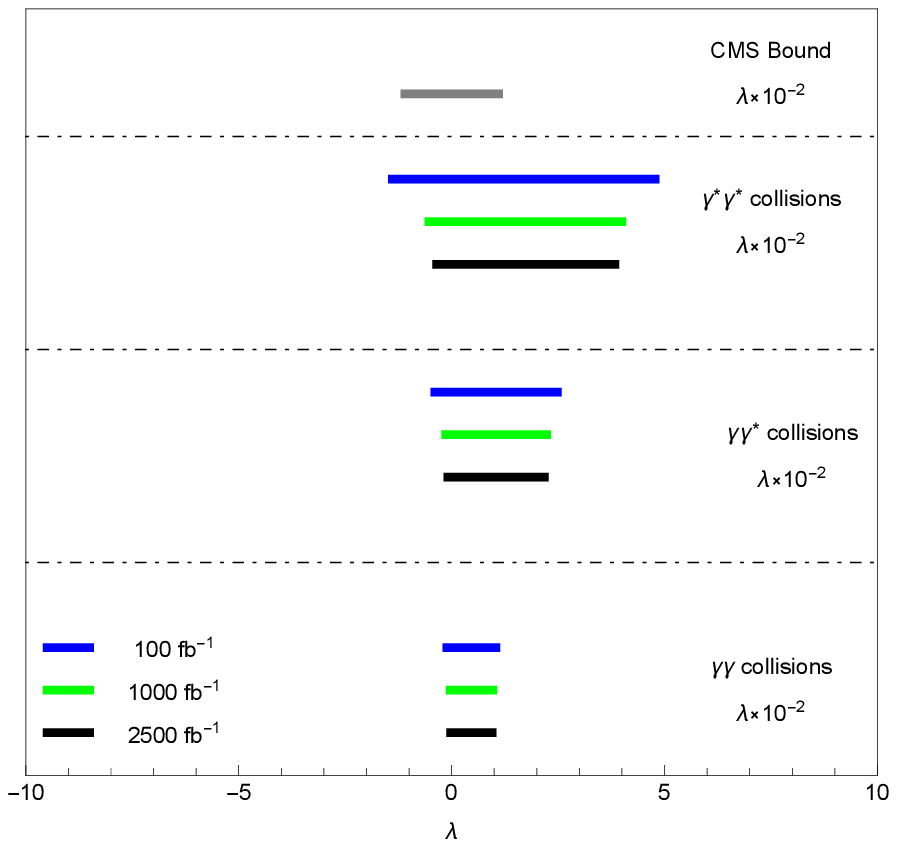}}}
\caption{ \label{fig:gamma4} Comparison of precisions at the CLIC to the anomalous coupling $\lambda_\gamma$
for center-of-mass energy $\sqrt{s}=1.5\hspace{0.8mm}{\rm TeV}$ and luminosities
${\cal L}=100, 1000, 2500\hspace{0.8mm}{\rm fb^{-1}}$. The figure covers the three processes
$\gamma\gamma\rightarrow W^+ W^-$, $e^+ \gamma \rightarrow e^+ \gamma^{*} \gamma \rightarrow e^+ W^- W^+$
and $e^+ e^-\rightarrow e^+ \gamma^{*} \gamma^{*} e^-\rightarrow e^+ W^- W^+ e^-$. We include the CMS bound.}
\end{figure}

\begin{figure}[H]
\centerline{\scalebox{1}{\includegraphics{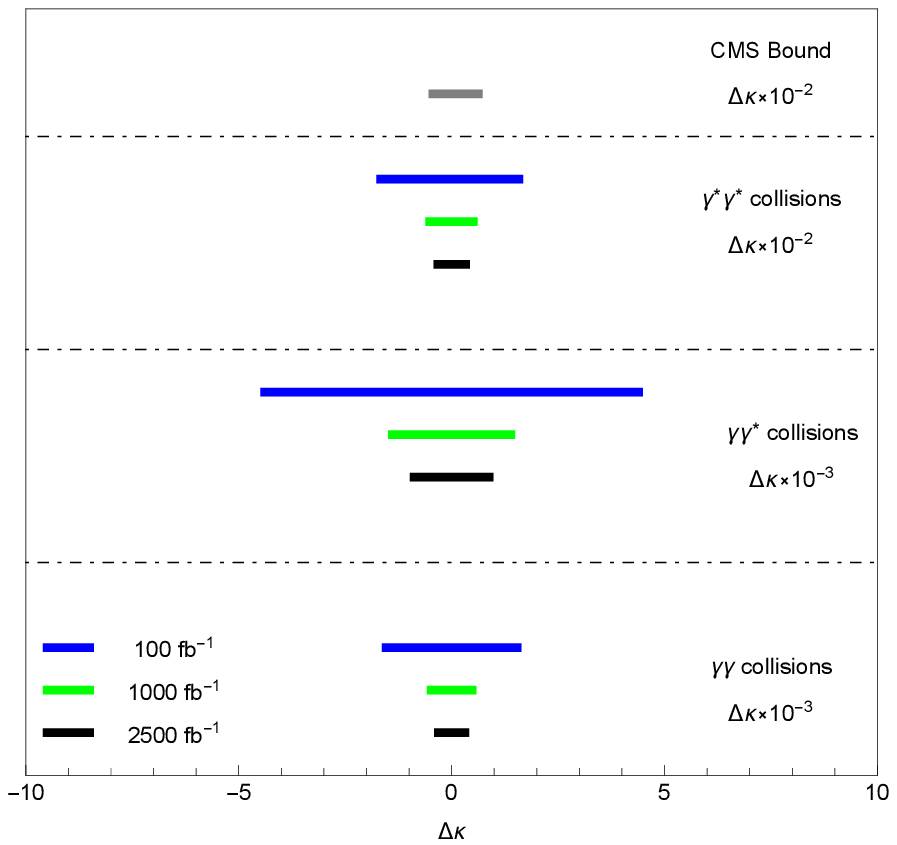}}}
\caption{ \label{fig:gamma15} The same as in Fig. 12, but for $\Delta\kappa_\gamma$.}
\end{figure}

\begin{figure}[H]
\centerline{\scalebox{1}{\includegraphics{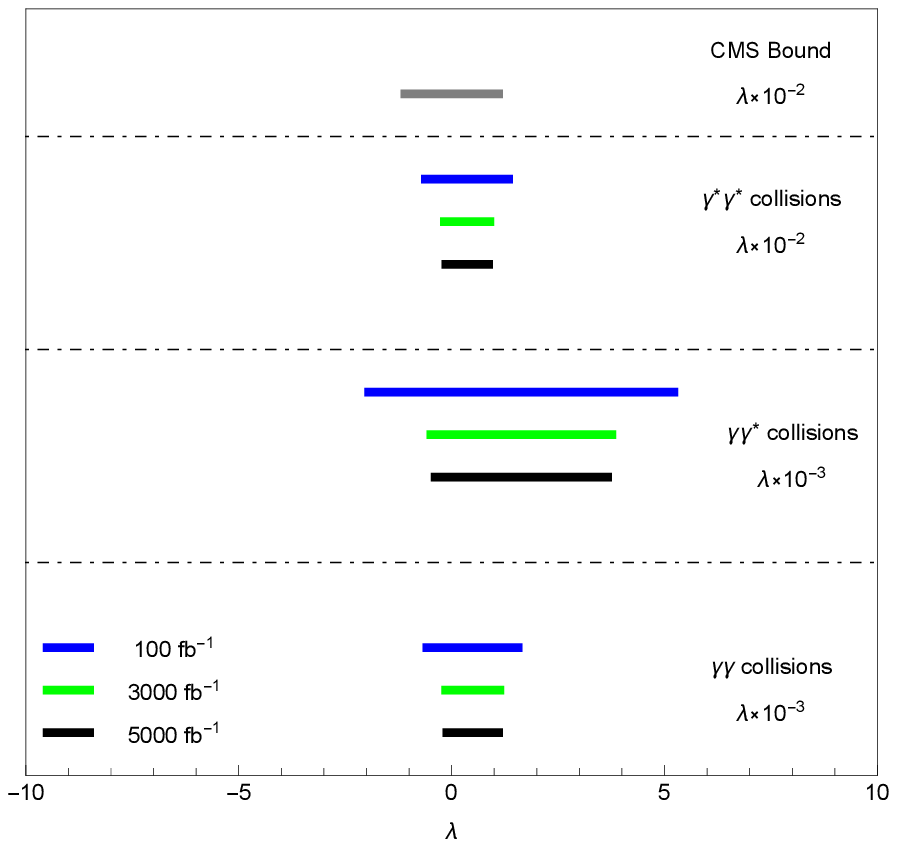}}}
\caption{\label{fig:gamma16} Comparison of precisions at the CLIC to the anomalous coupling $\lambda_\gamma$
for center-of-mass energy $\sqrt{s}=3\hspace{0.8mm}{\rm TeV}$ and luminosities
${\cal L}=100, 3000, 5000\hspace{0.8mm}{\rm fb^{-1}}$. The figure covers the three processes
$\gamma\gamma\rightarrow W^+ W^-$, $e^+ \gamma \rightarrow e^+ \gamma^{*} \gamma \rightarrow e^+ W^- W^+$
and $e^+ e^-\rightarrow e^+ \gamma^{*} \gamma^{*} e^-\rightarrow e^+ W^- W^+ e^-$. We include the CMS bound.}
\end{figure}

\begin{figure}[H]
\centerline{\scalebox{1}{\includegraphics{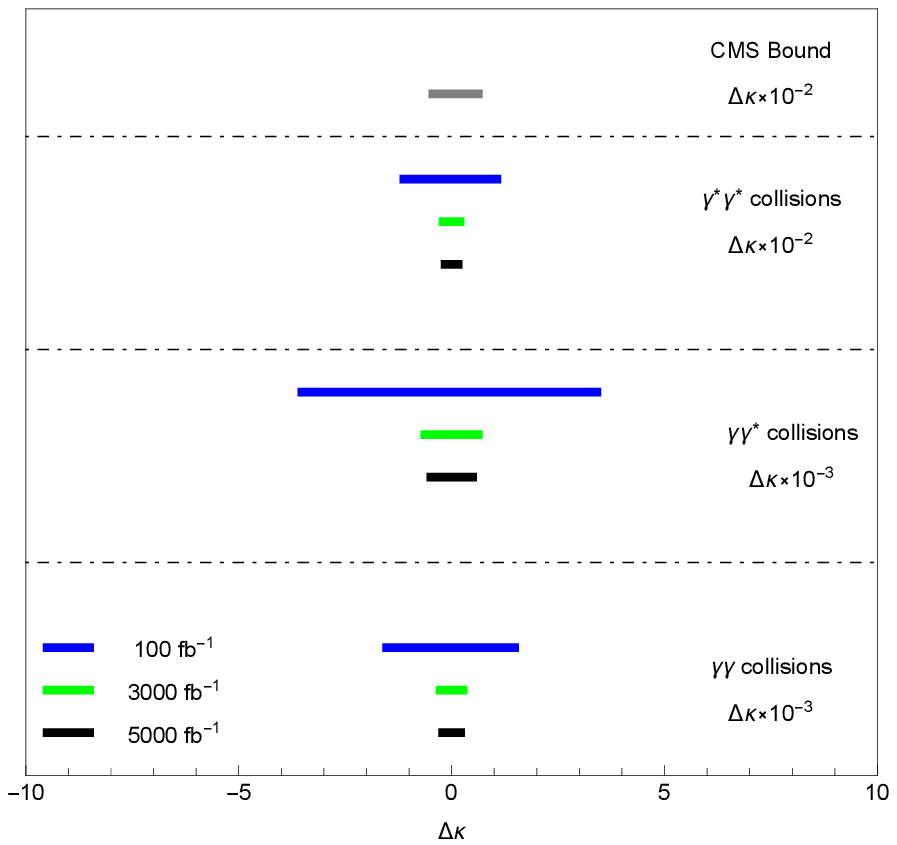}}}
\caption{\label{fig:gamma17} The same as in Fig. 14, but for $\Delta\kappa_\gamma$.}
\end{figure}

\begin{figure}[H]
\centerline{\scalebox{1}{\includegraphics{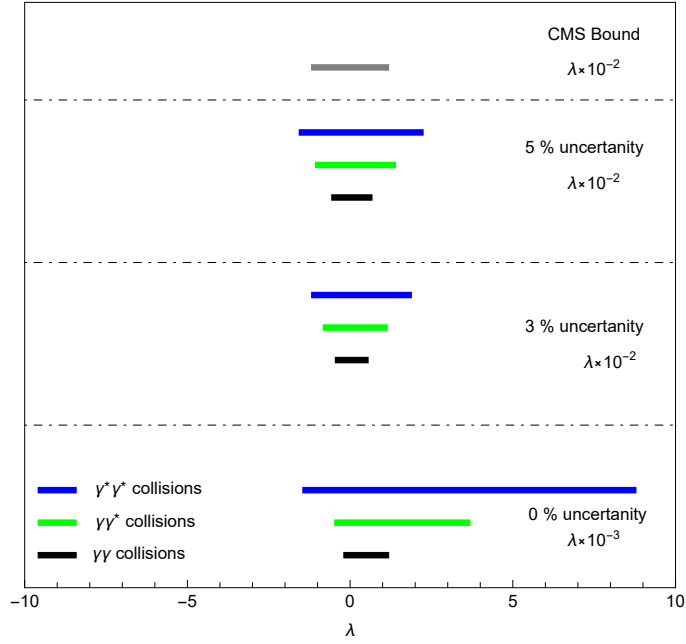}}}
\caption{ \label{fig:gamma5} Comparison of precisions at the CLIC to the anomalous coupling $\lambda_\gamma$
for center-of-mass energy $\sqrt{s}=3\hspace{0.8mm}{\rm TeV}$ and luminosity ${\cal L}=5000\hspace{0.8mm}{\rm fb^{-1}}$
with systematic uncertainties $\delta_{sys}=0, 3, 5\%$. The figure covers the three processes
$\gamma\gamma\rightarrow W^+ W^-$, $e^+ \gamma \rightarrow e^+ \gamma^{*} \gamma \rightarrow e^+ W^- W^+$ and
$e^+ e^-\rightarrow e^+ \gamma^{*} \gamma^{*} e^-\rightarrow e^+ W^- W^+ e^-$, with pure leptonic decays.
We include the CMS bound.}
\end{figure}

\begin{figure}[H]
\centerline{\scalebox{1}{\includegraphics{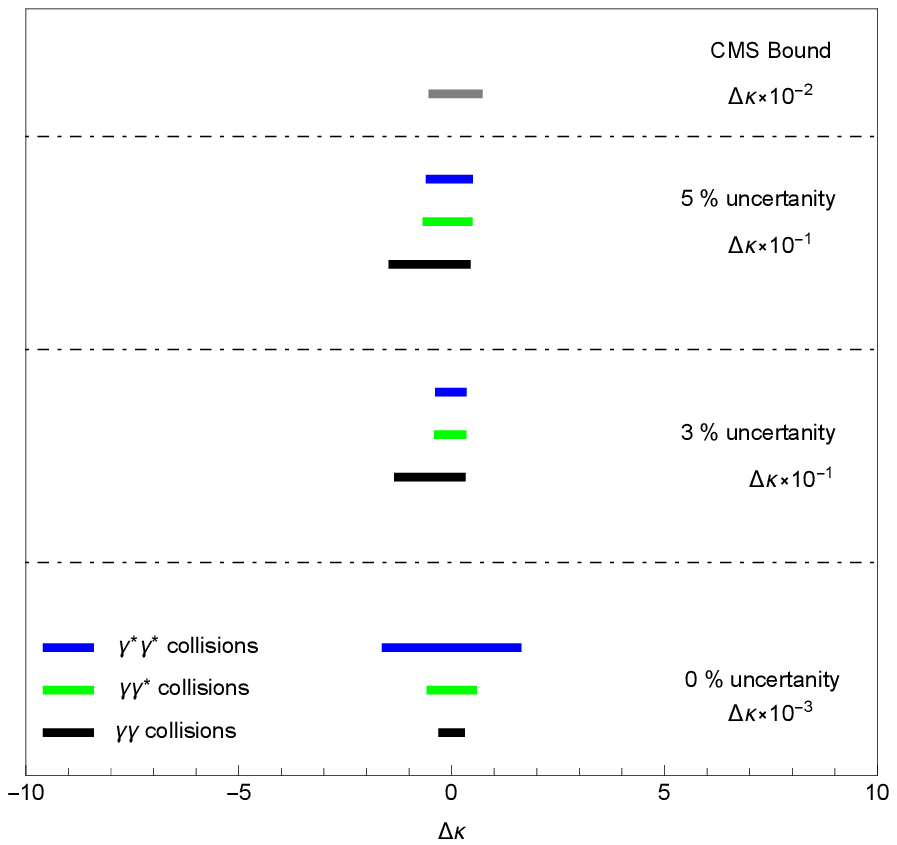}}}
\caption{\label{fig:gamma17} The same as in Fig. 16, but for $\Delta\kappa_\gamma$.}
\end{figure}

\end{document}